\documentclass[journal=nalefd,manuscript=letter]{achemso}

\usepackage[version=3]{mhchem} 
\usepackage{graphicx}
\usepackage{amsmath,amssymb}
\usepackage{siunitx}
\DeclareSIUnit{\rad}{rad}
\DeclareSIUnit{\oe}{Oe}
\usepackage{booktabs}
\usepackage{array}
\usepackage{cancel}

\newcommand{\Oe}{\ensuremath{\text{Oe}}}
\newcommand{\nm}{\ensuremath{\text{nm}}}
\newcommand{\nA}{\ensuremath{\text{nA}}}

\author{Muqing Yu}
\affiliation[University of Pittsburgh]
{Department of Physics and Astronomy, University of Pittsburgh, Pittsburgh, PA 15260, USA}

\author{Jieun Kim}
\affiliation[University of Wisconsin-Madison]
{Department of Materials Science and Engineering, University of Wisconsin-Madison, Madison, WI 53706, USA}

\author{Ahmed Omran}
\affiliation[University of Pittsburgh]
{Department of Physics and Astronomy, University of Pittsburgh, Pittsburgh, PA 15260, USA}

\author{Zhuan Li}
\affiliation[University of Pittsburgh]
{Department of Physics and Astronomy, University of Pittsburgh, Pittsburgh, PA 15260, USA}

\author{Jiangfeng Yang}
\affiliation[University of Wisconsin-Madison]
{Department of Materials Science and Engineering, University of Wisconsin-Madison, Madison, WI 53706, USA}

\author{Sayanwita Biswas}
\affiliation[University of Pittsburgh]
{Department of Physics and Astronomy, University of Pittsburgh, Pittsburgh, PA 15260, USA}

\author{Chang-Beom Eom}
\affiliation[University of Wisconsin-Madison]
{Department of Materials Science and Engineering, University of Wisconsin-Madison, Madison, WI 53706, USA}

\author{David Pekker}
\affiliation[University of Pittsburgh]
{Department of Physics and Astronomy, University of Pittsburgh, Pittsburgh, PA 15260, USA}

\author{Patrick Irvin}
\affiliation[University of Pittsburgh]
{Department of Physics and Astronomy, University of Pittsburgh, Pittsburgh, PA 15260, USA}

\author{Jeremy Levy}
\email{jlevy@pitt.edu}
\affiliation[University of Pittsburgh]
{Department of Physics and Astronomy, University of Pittsburgh, Pittsburgh, PA 15260, USA}

\title[KTaO\(_3\)-Based Supercurrent Diode]
  {KTaO\(_3\)-Based Supercurrent Diode}

\keywords{supercurrent diode effect, KTaO\(_3\), oxide interface, conductive AFM lithography, vortex dynamics}

\begin{document}

\begin{abstract}
The supercurrent diode effect (SDE), characterized by nonreciprocal critical currents, represents a promising building block for future dissipationless electronics and quantum circuits. Realizing SDE requires breaking both time-reversal and inversion symmetry in the device. Here we use conductive atomic force microscopy (c-AFM) lithography to pattern reconfigurable superconducting weak links (WLs) at the LaAlO\(_3\)/KTaO\(_3\) (LAO/KTO) interface. By deliberately engineering the WL geometry at the~nanoscale, we realize SDE in these devices in the presence of modest out-of-plane magnetic fields. The SDE polarity can be reversed by simply changing the WL position, and the rectification efficiency reaches up to 13\% under optimal magnetic field conditions. Time-dependent Ginzburg-Landau simulations reveal that the observed SDE originates from asymmetric vortex motion in the inversion-symmetry-breaking device~geometry. This demonstration of SDE in the LAO/KTO system establishes a versatile platform for investigating and engineering vortex dynamics, forming the basis for engineered quantum circuit elements.
\end{abstract}

\section{Introduction}
The supercurrent diode effect (SDE) refers to the non-reciprocal current flow in a superconductor, where its critical current differs significantly depending on the direction of the current. This asymmetry, analogous to semiconductor diodes, enables rectification of alternating currents in superconducting circuits and represents a useful component for low-dissipation quantum electronics. SDE has been reported in various material platforms \cite{Ando2020-eg,Pal2022-kk,Baumgartner2022-gu,Bauriedl2022-le,Wu2022-xj,Lin2022-mm,Zhao2023-vj,Diez-Merida2023-qn,Ghosh2024-yi,Zhang2025-rv}, with multiple theoretical explanations proposed \cite{Daido2022-mi,He2022-rd,Yuan2022-rv,Hou2023-if}. Although the exact mechanism varies between systems, two necessary ingredients are common to all reports. First, inversion symmetry must be broken either in the crystal structure or the device~geometry. Second, time-reversal symmetry must be broken by an external magnetic field \cite{Ando2020-eg,Pal2022-kk} or internally by spontaneous magnetic order \cite{Lin2022-mm,Zhao2023-vj}. The diode rectification efficiency $\eta$ is defined as $\eta=\frac{I_{c+}-|I_{c-}|}{I_{c+}+|I_{c-}|}$, where $I_{c+}$ and $I_{c-}$ are the critical currents in opposite directions. The sign and magnitude of $\eta$ indicate the polarity and strength of SDE, respectively. Realizing SDE with controllable polarity and high rectification efficiency remains an active area of research.

KTaO\(_3\) (KTO), specifically its heterointerface with LaAlO\(_3\) (LAO), has recently emerged as a platform for studying two-dimensional superconductivity \cite{Liu2021-eq, Chen2021-dj}. The 111-oriented LAO/ KTO interface exhibits superconductivity with $T_c$ up to 2 K (Ref. \citenum{Liu2021-eq}) while offering exceptional flexibility through conductive atomic force microscope (c-AFM) lithography. Nanoscale superconducting devices can be written and erased in a reconfigurable manner by sketching on the LAO/KTO surface with a biased AFM tip~\cite{Yu2022-pf,Hong2022-ch,Yu2025-ey,Wang2025-fo}. Superconducting weak links (WLs), essential components for superconducting circuits, were previously realized on LAO/KTO~\cite{Yu2025-ey}. In this work, we report SDE in c-AFM-patterned KTO WLs when time-reversal symmetry is broken by an external magnetic field and inversion symmetry is broken by deliberately displacing WL from the centerline of the device. We demonstrate control over both the polarity and magnitude of SDE by varying the WL position. The rectification efficiency $|\eta|$ reaches up to approximately 13\% under optimal magnetic field conditions. Previous studies have highlighted the critical role of vortex dynamics in the SDE of two-dimensional superconductors~\cite{Hou2023-if,Cerbu2013-ld,Suri2022-fa}. Through time-dependent Ginzburg-Landau simulations, we ascribe the origin of SDE to the combination of Meissner screening currents and asymmetric vortex surface barriers in the KTO WLs. Two of the studied WLs exhibit SDE combined with enhanced superconductivity at finite magnetic fields, where both $I_{c+}$ and $|I_{c-}|$ increase as the field deviates from zero. One possibility is that this effect is caused by the magnetization of local magnetic moments in the KTO sample~\cite{Kharitonov2005-qt,Wei2006-ov,Rogachev2006-vt}. Alternatively, this could be a signature of quantization of the number of vortices in the device, that is the Weber blockade~\cite{Pekker2011-yq,Morgan-Wall2015-qf}.

\section{Results}

\subsection{Supercurrent diode effect}
We report six WLs (Devices A through F) patterned by c-AFM lithography at the LAO/KTO (111) interface (see Methods for details). Devices B through F exhibit clear SDE under finite magnetic fields applied perpendicular to the Device~plane ($B=B_z$), while Device~A serves as a reference with suppressed SDE magnitude.

\begin{figure}
\includegraphics[scale=0.97]{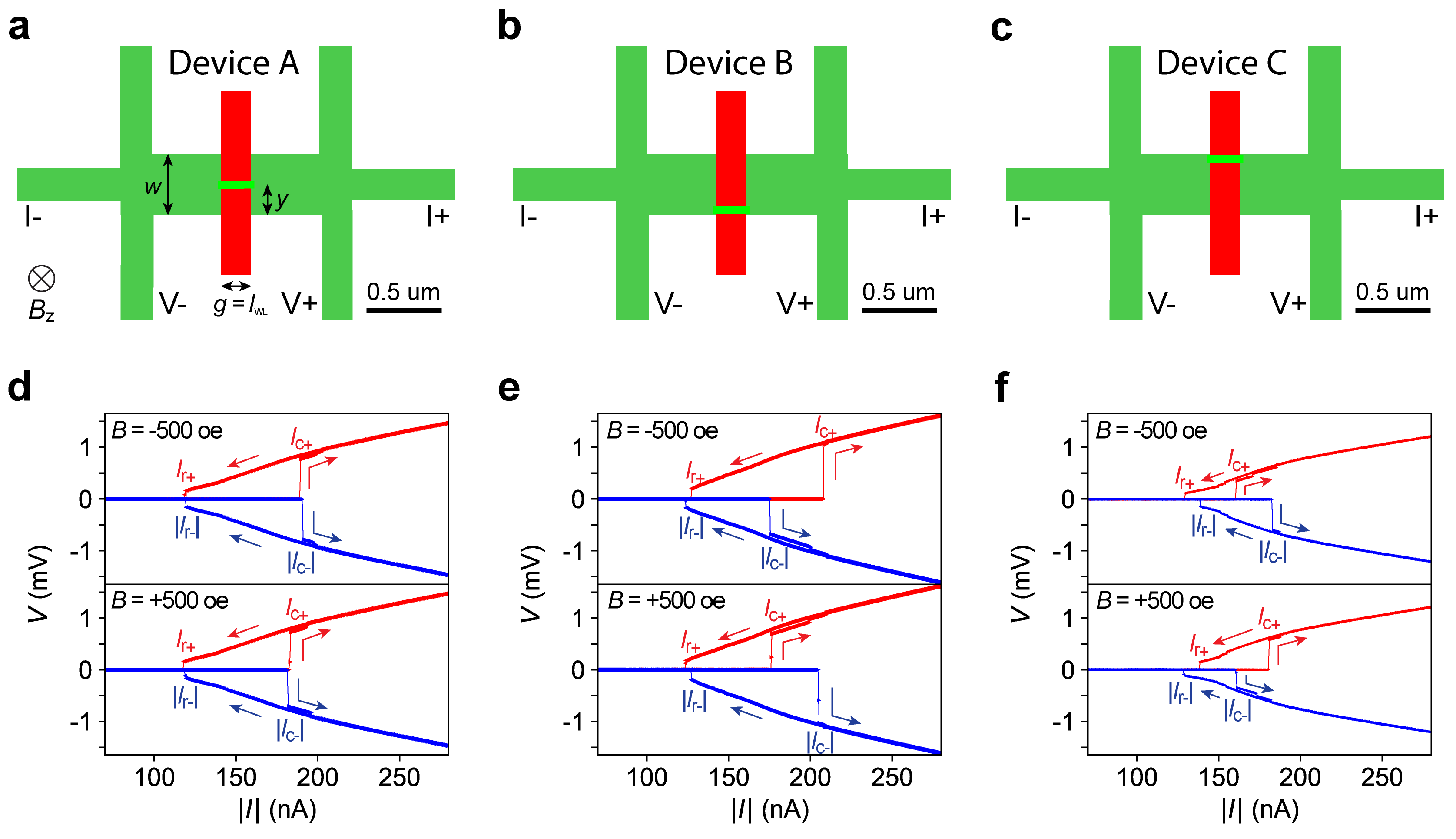}
\caption{Supercurrent diode effect in KTO WLs A-C. \textbf{(a)} Layout of the reference Device~A. Device~A is created by cutting a 2D channel (dark green, width $w=400\,\nm$) into the left and right halves by the red rectangle, and then bridging them with a nanowire (light green path) which serves as the WL. The gap $g$ created by the cutting corresponds with the length of the weak link: $l_{WL}=g$, which is estimated to be $\approx200\,\nm$ (see Supplementary Note~S2
). The WL is centered in the vertical direction. We define $y$ to be the vertical distance between the center of WL to the bottom edge of the 2D channel, which equals $w/2=200\,\nm$. \textit{I+}, \textit{I-}, \textit{V+} and \textit{V-} indicate the current source, current drain and the two voltages leads  used in the following four-terminal $I-V$ measurements. Positive bias current ($I>0$) flows from the right to the left. Positive magnetic field ($B>0$) points into the sample plane. (b) Layout of Device~B, where WL is placed close to the bottom edge of the 2D channel ($w=400\,\nm$, $y=32\,\nm$). (c) Layout of Device~C, where WL is placed close to the top edge of the 2D channel ($w=400\,\nm$, $y=368\,\nm$). (d) $I-V$ measurements of Device~A at $B=-500\,\Oe$ (top) and $B=+500\,\Oe$ (bottom). The red curve is the $V$ vs $I$ curve under positive current ($I>0$) while the blue curve is $V$ vs $|I|$ curve under negative current ($I<0$). The arrows indicate the current sweep directions, while switching currents $I_{c+}$, $|I_{c-}|$ and retrapping currents $I_{r+}$, $|I_{r-}|$ are labeled. (e) $I-V$ measurements of Device~B, where obvious mismatch between $I_{c+}$ and $|I_{c-}|$ can be observed. At $B=-500\,\Oe$, $I_{c+}>|I_{c-}|$ while at $B=+500\,\Oe$, $I_{c+}<|I_{c-}|$. (f) $I-V$ measurements of Device~C. At $B=-500\,\Oe$, $I_{c+}<|I_{c-}|$ while at $B=+500\,\Oe$, $I_{c+}>|I_{c-}|$. Note: all plots in this figure were taken at $T=50$~mK with a backgate voltage $V_{bg}=-30$~V applied on Devices A-C.}
\label{ABC IV}
\end{figure}

Device~A (Figure~\ref{ABC IV}(a)) exemplifies an inversion-symmetric WL geometry. A horizontal 2D superconducting channel (width $w=400\,\nm$) is divided into left and right halves and then bridged by a superconducting~nanowire (WL) at the center. The WL is positioned at the vertical center of the channel, equidistant from both edges. Details of the c-AFM lithography process are provided in Ref.~\citenum{Yu2025-ey} and Methods. During cryogenic measurements at $T=50$~mK, a magnetic field perpendicular to the sample plane is applied. Current-voltage ($I$-$V$) measurements of Device~A at $B=-500\,\Oe$ and $B=+500\,\Oe$ are shown in the top and bottom panels of Figure~\ref{ABC IV}(d), respectively. At both field values, $V$ remains at zero as $I$ increases from zero until an abrupt transition to the normal state occurs at the positive critical current (or positive switching current) $I_{c+}$. As $I$ decreases from its positive maximum back to zero, the device returns to the superconducting state at the positive retrapping current $I_{r+}$ (red curves, Figure~\ref{ABC IV}(d)). Similarly, the negative critical current $I_{c-}$ and negative retrapping current $I_{r-}$ are observed as $I$ sweeps from zero in the negative direction and back (blue curves, Figure~\ref{ABC IV}(d)). The hysteretic $I$-$V$ characteristics indicate that the WL operates in the underdamped regime or experiences self-heating in the normal state. At $B=\pm 500\,\Oe$, Device~A exhibits minimal diode rectification,as $I_{c+}$ and $|I_{c-}|$ differs by $<3\,\nA$ which corresponds to $|\eta|<1\%$.

The values of $I_{c+}$ and $|I_{c-}|$ deviate from each other when inversion symmetry is deliberately broken in the device~layout by displacing the WL to on one side of the channel. Device~B is patterned identically to Device~A except that the WL is positioned near the bottom edge of the channel (distance from the bottom: $y=32\,\nm$, Figure~\ref{ABC IV}(b)). Under a negative magnetic field $B=-500\,\Oe$, Device~B exhibits pronounced SDE with $I_{c+}$ exceeding $|I_{c-}|$ by approximately 30~nA ($\eta=+8.3\%$, Figure~\ref{ABC IV}(e) top panel). When $B$ switches to $+500\,\Oe$, the SDE polarity in Device~B changes sign, as $I_{c+}$ now falls below $|I_{c-}|$ by approximately 30~nA ($\eta=-8.0\%$, Figure~\ref{ABC IV}(e) bottom panel). Remarkably, the SDE polarity can be flipped by ``flipping" the WL position. In Device~C, the position of the WL is shifted from bottom to top of the 2D channel ($y=368\,\nm$, Figure~\ref{ABC IV}(c)) relative to Device~B. This results in $I_{c+}<|I_{c-}|$ at $B=-500\,\Oe$ ($\eta=-6.4\%$, Figure~\ref{ABC IV}(f) top panel) while $I_{c+}>|I_{c-}|$ at $B=+500\,\Oe$ ($\eta=+6.5\%$, Figure~\ref{ABC IV}(f) bottom panel), opposite to the behavior of Device~B.

\begin{figure}
\includegraphics[scale=0.99]{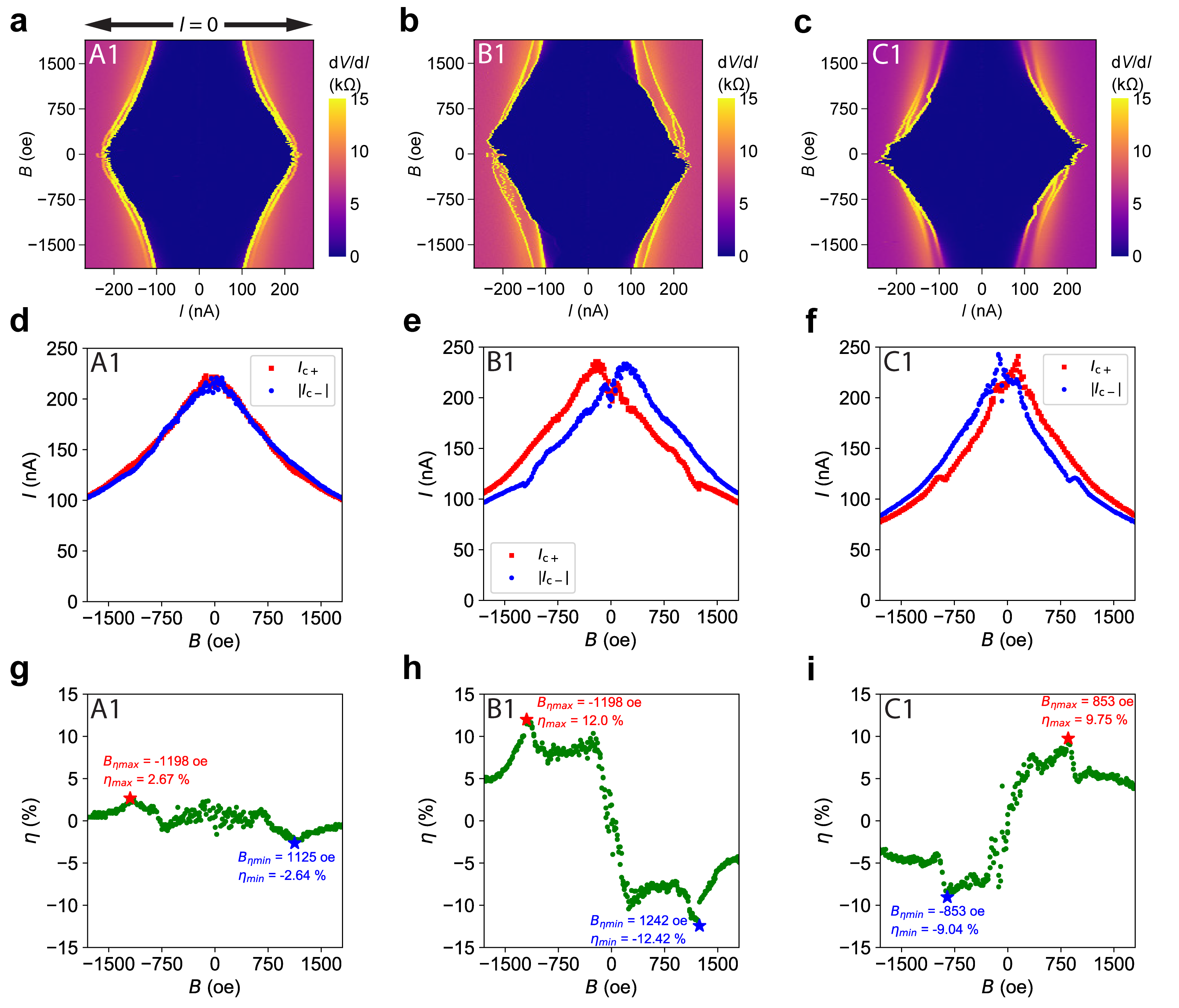}
\caption{Magnetic field sweep of Devices A through C. Panels (a), (b), and (c) show intensity plots of differential resistance $dV/dI$ versus $I$ versus $B$ for Devices A, B, and C, respectively. We note that in these plots, current $I$ sweeps from $I=0$ to $|I|>0$ to capture the switching behavior from superconducting state to normal state. Panels (d), (e), and (f) display the extracted switching currents $I_{c\pm}$ as a function of $B$ for Devices A, B, and C. Panels (g), (h), and (i) show the extracted diode efficiency $\eta$ as a function of $B$ for Devices A through C. On the top left corner of each panel, the label consists of a letter that indicates the corresponding Device, and a number that points to the measurement configuration (mapping in Figure~\ref{config}). All measurements were performed at $T=50$~mK with a backgate voltage $V_{\rm bg}=-30$~V.}
\label{ABC IcB}
\end{figure}

Continuous magnetic field sweeps provide additional information on how $I_{c\pm}$ and SDE strength evolve with $B$. Figure~\ref{ABC IcB}(a) through (c) presents intensity plots of $dV/dI$ as a function of $I$ and $B$ for Devices A through C, respectively. These plots take the portion of $I-V$ curves where magnitude of current $|I|$ increases from 0. This enables us to visualize and extract $I_{c\pm}$ at the points where $dV/dI$ increases above $R_N/2$ (half of the normal state resistance). We note that in all the following $dV/dI$ intensity plots, $|I|$ increases from 0 if not specifically labeled. The $dV/dI$ pattern of Device~A (reference device) appears symmetric with respect to $I=0$ (Figure~\ref{ABC IcB}(a)), with $I_{c+}$ and $|I_{c-}|$ nearly overlapping across the entire measured field range (Figure~\ref{ABC IcB}(d)). In contrast, the $dV/dI$ pattern of Device~B appears skewed (Figure~\ref{ABC IcB}(b)), with clear deviation between the two critical currents: $I_{c+}<|I_{c-}|$ at $B>0$ and $I_{c+}>|I_{c-}|$ at $B<0$ (Figure~\ref{ABC IcB}(e)). Despite this skewness, the $dV/dI$ and $I_c$ of Device~B follow inversion symmetry with respect to field $B$ and bias $I$:
\begin{equation}
\label{inv2}
dV/dI|_{I,B}=dV/dI|_{-I,-B}
\end{equation}
\begin{equation}
\label{inv}
I_{c+}(B)=-I_{c-}(-B)
\end{equation}
These relationships are theoretically expected and experimentally observed in systems without intrinsic time-reversal symmetry breaking~\cite{Zhang2025-rv}. As shown in Figure~\ref{ABC IcB}(h), $\eta$ for Device~B transitions from positive to negative approximately linearly with $B$ as the field increases from $-200$~Oe to $+200$~Oe. The efficiency reaches its maximum $\eta_{\rm max}=+12.0\%$ at the optimal field $B_{\eta \rm max}=-1198\,\Oe$ and its minimum $\eta_{\rm min}=-12.4\%$ at $B_{\eta \rm min}=+1242\,\Oe$ (Figure~\ref{ABC IcB}(h)). At $|B|>1500\,\Oe$, the difference between $I_{c+}$ and $|I_{c-}|$ is suppressed, reducing $|\eta|$ accordingly. We note that $\eta_{\rm max}\approx -\eta_{\rm min}$ and $B_{\eta \rm max}\approx -B_{\eta \rm min}$ due to the symmetry expressed in Eq.~\ref{inv}. For Device~C, with the WL positioned on the opposite edge of the channel, the $dV/dI$ versus $I$ versus $B$ pattern (Figure~\ref{ABC IcB}(c)) and critical currents $I_{c\pm}(B)$ (Figure~\ref{ABC IcB}(f)) exhibit opposite skew compared to Device~B. Device~C achieves $\eta_{\rm max}=+9.75\%$ at $B_{\eta \rm max}=+853\,\Oe$ and $\eta_{\rm min}=-9.04\%$ at $B_{\eta \rm min}=-853\,\Oe$ (Figure~\ref{ABC IcB}(i)). In Device~B and C, there is also slight mismatch between the positive and negative retrapping currents (Figure~\ref{ABC IrB}), but less significant than the difference between $I_{c+}(B)$ and $I_{c-}(B)$. For the reference Device~A, with its inversion-symmetric layout, $\eta(B)$ remains confined within $\pm3\%$ across the entire field range (Figure~\ref{ABC IcB}(g)).

Comparison among Devices A through C reveals two key findings. First, strong SDE in KTO WLs requires breaking both time-reversal symmetry through the applied field $B$ and inversion symmetry through the device~geometry. Second, the sign of $\eta$ can be controlled by varying the WL position. Electrostatic gating is applied on Device~A, B and C by a voltage $V_{bg}$ on the backside of the sample \cite{Fujii1976-ph,Chen2021-dj}. The effect of $V_{bg}$ on SDE is discussed in Supplementary Note S1.
\ Device~F, created using an alternative c-AFM process in which the channel is only partially cut to leave a thin conducting path near the bottom edge (Figure~\ref{F}), also demonstrates SDE with the same polarity as Device~B, providing an alternative approach to KTO supercurrent diode patterning.

\subsection{SDE combined with field enhancement of superconductivity}
\begin{figure}
\includegraphics[scale=0.95]{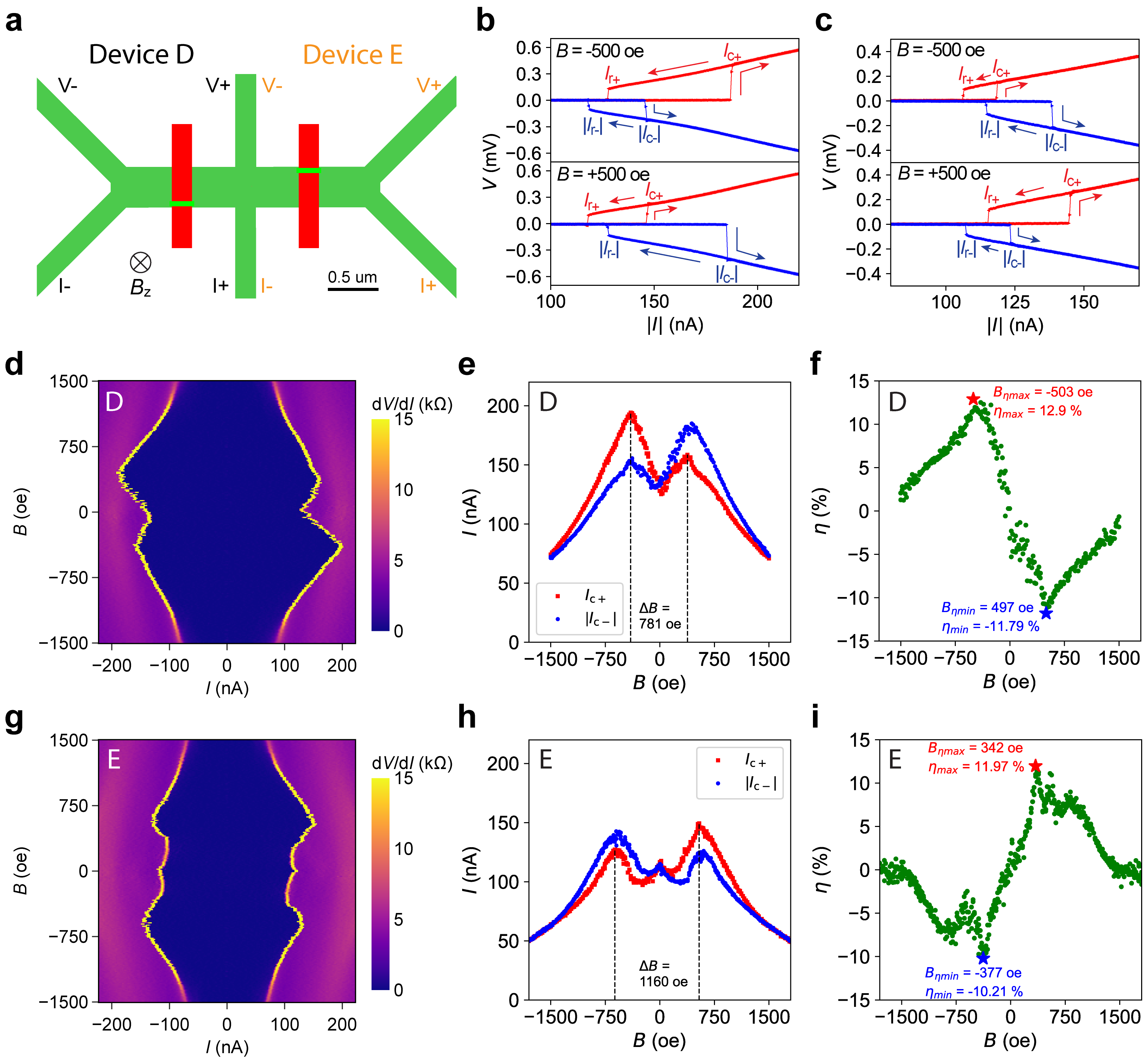}
\caption{Supercurrent diode D and E with slanted M-shaped $I_c$ vs $B$ pattern. (a) Device~layout. Device~D and E were patterned together in one run by c-AFM lithography. Both WLs are put in the $w=400\,\nm$ 2D channel with WL D(E) positioned at $y=36\,\nm$ ($y=364\,\nm$). The measurement configuration for Device~D(E) is indicated in black(orange) $I+/I-/V+/V-$ labels, as two leads are shared between them. (b)(c) $I$-$V$ measurements of Device~D(E) at $B=\pm500\,\Oe$. (d) $dV/dI$ vs $I$ vs $B$ intensity plot of Device~D. (e) $I_{c\pm}$ vs $B$ of Device~D, which follows a slanted ``M'' pattern. Black dashed lines label the two $I_{c+}$ maxima at $B=+400\,\Oe$ and at $B=-380\,\Oe$, while $I_{c+}(B=0)$ is lower than either of these maxima. (f) $\eta$ vs $B$ relation of Device~D. (g)(h)(i) $dV/dI$ vs $I$ vs $B$ intensity plot, $I_{c\pm}$ vs $B$ and $\eta$ vs $B$ relations of Device~E. In panel (h), the two $I_{c+}$ maxima occur at $B=+540\,\Oe$ and at $B=-620\,\Oe$. All plots in this figure taken at $T=50$~mK with backgate grounded $V_{\rm bg}=0$~V on Devices D \& E.}
\label{DE IcB}
\end{figure}

Two additional supercurrent diodes, Devices D and E, were created by c-AFM lithography. As shown in Figure~\ref{DE IcB}(a), inversion symmetry is again broken in these devices by positioning the WL near the bottom edge in Device~D ($y=36\,\nm$) and near the top edge in Device~E ($y=364\,\nm$). Their $I$-$V$ characteristics at $T=50$~mK and $B=\pm500\,\Oe$ again show asymmetry between positive and negative critical currents (Figure~\ref{DE IcB}(b) and (c)), with hysteretic behavior and distinct retrapping currents (Figure~\ref{DE IrB}). The SDE polarity in Device~D matches that of Device~B, with both exhibiting $I_{c+}>|I_{c-}|$ at $B<0$ and $I_{c+}<|I_{c-}|$ at $B>0$ (Figure~\ref{DE IcB}(e)). Conversely, Devices C and E show consistent behavior with $I_{c+}<|I_{c-}|$ at $B<0$ and $I_{c+}>|I_{c-}|$ at $B>0$ (Figure~\ref{DE IcB}(h)). This further confirms that WL position is the decisive factor determining SDE polarity. The efficiency $\eta$ of Devices D and E reaches $\eta_{\rm max}>+10\%$ and $\eta_{\rm min}<-10\%$ (Figure~\ref{DE IcB}(f) and (i)), comparable to the $\eta_{max(min)}$ measured in Devices B and C.

Beyond this asymmetry, the field-dependent evolution of $I_{c+}(B)$ and $I_{c-}(B)$ in these devices warrants attention. In Device~D, rather than the typical suppression of $I_c$ by magnetic field, $I_{c+}$ increases from 135~nA at $B=0$ to 160~nA at $B=+380\,\Oe$ (red data points, Figure~\ref{DE IcB}(e)). This constitutes a local $I_{c+}$ peak on the $B>0$ side, after which $I_{c+}$ decreases with further field increase. When $B$ decreases from zero in the negative direction, $I_{c+}$ again increases from 135~nA, reaching $I_{c+}=190\,\nA$ at $B=-400\,\Oe$ before decreasing. We define the field separation between the two $I_{c+}$ peaks as $\Delta B=780\,\Oe$. Similar enhancement of $I_c$ at finite field occurs in Device~E, whose $I_{c+}(B)$ exhibits two peaks away from $B=0$. The slanted M-shaped $I_c(B)$ pattern in Devices D and E results from the combination of two effects. First, the slant originates from SDE. Second, the M-shape reflects enhanced $I_c$ by magnetic field, and we discuss its possible origin in the Discussion Section
.

\section{Discussion}

\subsection{Origin of the supercurrent diode effect in KTO WLs}
\begin{figure}
\includegraphics[scale=0.97]{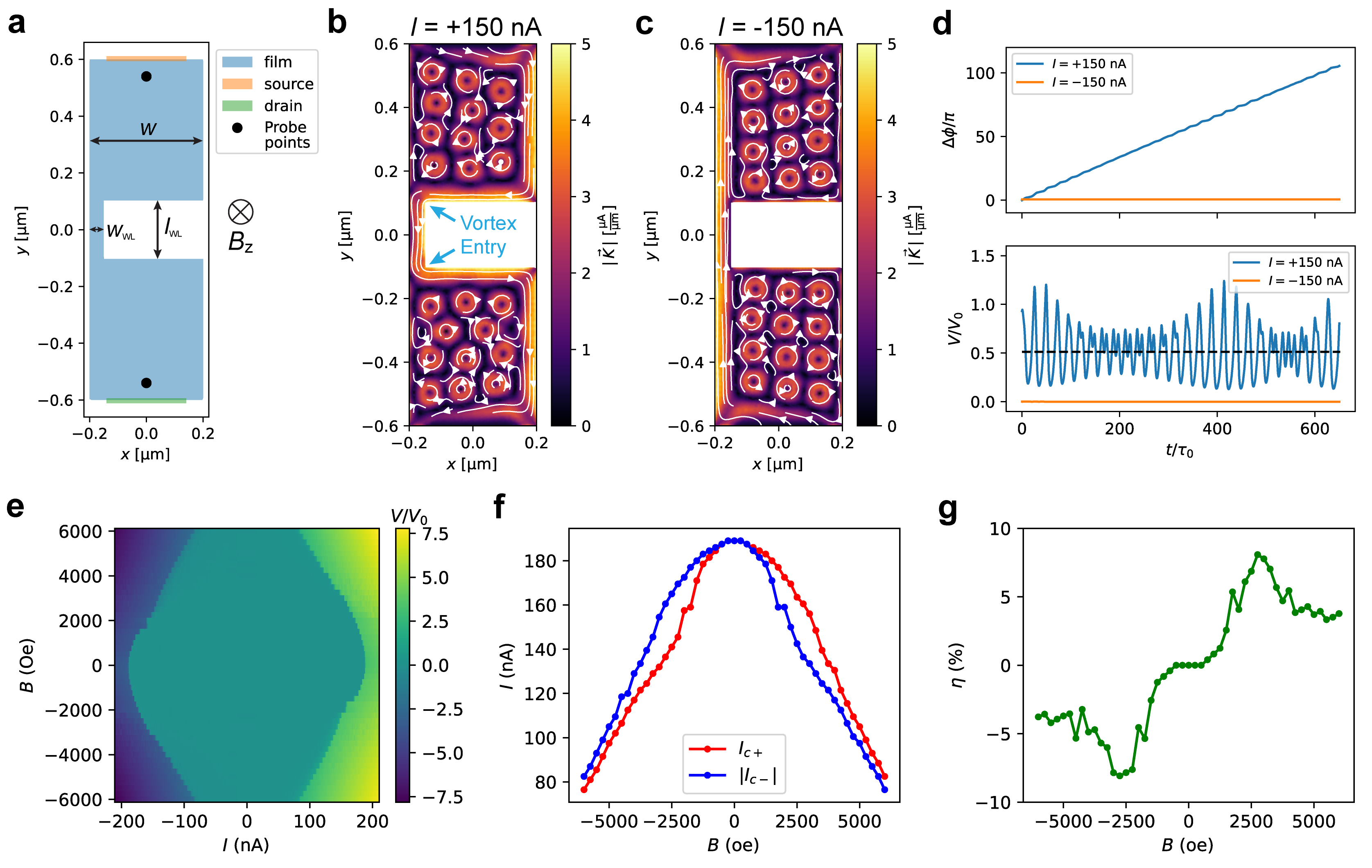}
\caption{Simulation of KTO WL using time-dependent Ginzburg-Landau theory. (a) Device geometry used in TDGL simulation. The 2D channel (blue) has width $w=400\,\nm$ and length $l=1200\,\nm$. On its top and bottom edges there are current source and drain, indicated by the orange and green bars respectively. Positive current $I>0$ flows from the top to the bottom. The narrow constriction (WL) has width $w_{WL}=50\,\nm$ and length $l_{WL}=200\,\nm$, located near the left edge of the 2D channel. Phase difference and voltage between the two black dots are output by TDGL calculation. We define positive magnetic field ($B>0$) to be pointing into the sample plane, same as the experimental setup. (b) Current density $\textbf{\textit{K}}(x,y)$ calculated under the condition $B=-2000\,\Oe$ and $I=+150\,\nA$. Color scale indicates magnitude of $\textbf{\textit{K}}$ while white arrows indicate its direction. The two blue arrows point to the two locations with relatively low surface barrier for vortex entry. (c) $\textbf{\textit{K}}(x,y)$ calculated at $B=-2000\,\Oe$ and $I=-150\,\nA$. (d) Evolution of phase difference $\Delta \phi$ and voltage $V$ as a function of time. Black dashed line: time-averaged voltage under $I=+150\,\nA$ bias. (e) $V$ vs $I$ vs $B$ intensity plot from pyTDGL simulation. (f) $I_{c\pm}$ vs $B$ relation extracted from the simulated $I$-$V$ curves. (g) $\eta$ vs $B$ relation extracted from (f).}
\label{tdgl}
\end{figure}

The observed SDE in KTO WLs exhibits two characteristic behaviors: the effect reverses sign upon reversing the out-of-plane magnetic field and upon repositioning the WL from one edge to the other. These observations suggest that Meissner currents play a central role in the SDE mechanism. Previous studies of SDE in thin metallic superconducting films have highlighted the importance of Meissner currents and provide a framework for our analysis~\cite{Cerbu2013-ld,Hou2023-if}. In two-dimensional superconducting systems, dissipation typically arises from vortex motion rather than Cooper pair breaking. To visualize vortex dynamics in our KTO WL geometry, we perform time-dependent Ginzburg-Landau (TDGL) simulations (see Methods and Ref.\cite{Bishop-Van_Horn2023-vs}). The simulated device~consists of a vertical channel ($w=400\,\nm$) with a narrow constriction near its left edge, constituting the WL (Figure~\ref{tdgl}(a)). Current source(drain) is defined at the top(bottom) edge of the channel. The simulated WL dimensions are $l_{\rm WL}=200\,\nm$ and $w_{\rm WL}=50\,\nm$, which are estimated based on Device~F characteristics, as detailed in Supplementary Note~S2.
\ By design, it closely resembles Device~C (Figure~\ref{ABC IV}(c)) rotated counterclockwise by $90^{\circ}$.

Figure~\ref{tdgl}(b) and (c) show the calculated current density $\textbf{\textit{K}}(x,y)$ under $B=-2000\,\Oe$ with applied currents of $I=+150\,\nA$ and $I=-150\,\nA$, respectively. Upon introduction of $B=-2000\,\Oe$, the Meissner effect induces a clockwise screening current. Also, some 11-12 vortices are introduced in the top and bottom halves of the device, represented by where \textbf{\textit{K}} forms circles. These vortices are static, not contributing to dissipation (see Supplementary Video 1 \& 2). The total current density is the sum of the applied current and the clockwise screening current, resulting in enhanced (suppressed) $|\textbf{\textit{K}}|$ along the right (left) edge of the device~under $I=+150\,\nA$ bias (Figure~\ref{tdgl}(b)). With supercurrent concentrating at the right edge of the device, the two corners near the WL (marked by arrows in Figure~\ref{tdgl}(b)) act as gateways for vortex entry due to the low surface barrier at these highly curved regions. After entry, these mobile vortices traverse the WL and exit at the left edge (Supplementary Video 1). Each vortex traveling from right to left causes the phase difference $\Delta \phi$ between the top and bottom of the device~to evolve by $2\pi$, corresponding to a voltage peak (blue curves, Figure~\ref{tdgl}(d)). The time-averaged voltage is indicated by the black dashed line in Figure~\ref{tdgl}(d), which would be the measurable DC voltage in experiments.

In contrast, a negative bias combined with the clockwise screening current produces enhanced(weakened) current density on the left(right) side of the Device~(Figure~\ref{tdgl}(c)). In this scenario, supercurrent concentrates at the left edge, which is relatively flat and thus presents a higher energy barrier for vortex entry. At $I=-150\,\nA$ throughout the simulated time frame, no mobile vortices enter the device (Supplementary Video 2), preventing phase slips and maintaining $V=0$ (orange curves, Figure~\ref{tdgl}(d)). The asymmetric vortex surface barriers at the two edges, combined with Meissner currents, lead to dissipation onset under $I=+150\,\nA$ but not under $I=-150\,\nA$. By performing TDGL calculations across a range of bias currents $I$ and magnetic fields $B$ and extracting the time-averaged DC voltage at each point (see Methods), we obtain the skewed $V$ versus $I$ versus $B$ pattern in Figure~\ref{tdgl}(e). We note the simulated $I_{c\pm}$ exhibit $I_{c+}(B)<|I_{c-}(B)|$ for $B<0$ and $I_{c+}(B)>|I_{c-}(B)|$ for $B>0$ (Figure~\ref{tdgl}(f)), consistent with the SDE polarity of Device~C. At $B=0$, SDE has the sign reversal due to the direction change of Meissner current under $B>0$ vs $B<0$. The simulated $I_{c\pm}(B)$ also follows the inversion symmetry expressed in Eq.~\ref{inv}. Regarding efficiency, the calculated $\eta$ reaches its maximum and minimum of $\eta_{\rm max}=-\eta_{\rm min}=+8.1\%$ at fields $B_{\eta \rm max}=-B_{\eta \rm min}=+2750\,\Oe$ (Figure~\ref{tdgl}(g)). The experimentally measured $\eta_{\rm max}$ and $|\eta_{\rm min}|$ of Devices B through F range between 9\% and 13\%, very close to the calculated value. Meanwhile the measured $B_{\eta max}$ and $|B_{\eta min}|$ are scattered, ranging from $300\,\Oe$ to $1300\,\Oe$, lower than the calculation results, which can be attributed to errors in the material parameters used in the simulation (see Methods). Inhomogeneities or defects within the KTO sample and fluctuations during c-AFM lithography can introduce random asymmetries and disorder such as rough edges and vortex pinning centers, which can also cause the discrepancy between measured and simulated $B_{\eta max(min)}$. The same randomly occurring asymmetries lead to the weakened but non-zero SDE in the reference Device~A ($|\eta|<3\%$, Figure~\ref{ABC IcB}(d)). The TDGL simulation not only highlights two essential ingredients for SDE (Meissner currents and asymmetric vortex surface barriers at the two edges \cite{Cerbu2013-ld,Hou2023-if}) but also demonstrates approximate quantitative agreement with the observed SDE strength in our KTO WLs.

During measurements of Devices A through C, we observed that the $I_{c\pm}(B)$ patterns depend on the choice of measurement configuration. We attribute this to the alternation of current profile and vortex surface barriers upon changing the position of current leads, which is verified qualitatively by TDGL simulation in Supplementary Note~S3.

\subsection{Origin of the magnetic field enhancement of $I_c$ in Devices D and E}
\label{M dis}
In Devices D and E, $I_c(B)$ exhibits two peaks (separated by $\Delta B$) where $I_c$ exceeds its zero-field value, roughly following a slanted M-shape. Previous studies reported enhancement of superconductivity by magnetic fields in Zn, MoGe, and Nb superconducting~nanowires \cite{Tian2005-xj, Rogachev2006-vt}. Specifically, Ref.~\citenum{Rogachev2006-vt} demonstrated an M-shaped $I_c(B)$ relation bearing great similarity to our observations, which was attributed to spin-exchange scattering between Cooper pairs and magnetic moments on the~nanowire surfaces. We believe a similar mechanism can qualitatively explain our observations in KTO WLs. At $B=0$, local magnetic moments in the KTO sample induce exchange scattering of electrons, effectively breaking Cooper pairs and weakening superconductivity \cite{Kharitonov2005-qt,Rogachev2006-vt,Wei2006-ov}. Consequently, vortex entry and nucleation become more energetically favorable, suppressing $I_c$. At finite $B$, the magnetic moments are aligned by the field and exchange scattering is quenched, leading to higher $I_c$. At high $B$, conventional orbital and Zeeman effects dominate, weakening superconductivity and causing $I_c$ to decrease. Competition between magnetic moment pair breaking and orbital and Zeeman pair breaking can produce a non-monotonic, M-shaped $I_c(B)$ curve. Superimposed with SDE, this yields the slanted M-shaped $I_c(B)$ curves observed.

The fact that magnetic moments are distributed unevenly in the KTO sample explains the absence of M-shape in $I_c(B)$ of Devices A-C. Several reports \cite{Krantz2024-iy,Ning2024-uq} have suggested magnetism in the LAO/KTO system, which is ascribed to Ta 5\textit{d} states and oxygen vacancies. Magnetic impurities such as Fe and Ni can also be brought into the sample, especially during the polishing process of the crystal \cite{Coey2016-zt}. Electrostatic gating with backgate ($V_{bg}$) is performed on Device~D and E (Figure~\ref{D bg} and Figure~\ref{E bg}). The magnitude of SDE $\eta_{max(min)}$ is barely affected by $V_{bg}$, always maxing out at $\approx10\%$ (top panel, Figure~\ref{DE bg}), similar to the case of Devices A-C (discussed in Supplementary Note S1
). Meanwhile, $\Delta B$ that separates the two $I_c$ peaks increases monotonically as $V_{bg}$ decreases (bottom panel, Figure~\ref{DE bg}), which suggests exchange scattering may be more pronounced with lower superfluid density and higher disorder. Enhancement of $I_c$ by field persists at temperature $T=500$~mK (Device~D, Figure~\ref{D temp}(b)). At $T=900$~mK however, $I_c(B)$ loses its M shape and only SDE can be seen (Figure~\ref{D temp}(e)). Further increase in $T$ greatly suppresses superconductivity and smears out the SDE (Figure~\ref{D temp}(h)). This temperature dependence indicates the onset of exchange scattering may occur at a temperature $\in(500\,\mathrm{mK}, 900\,\mathrm{mK})$, lower than the superconducting critical temperature $T_c$.

Another possible origin for the non-monotonic $I_c(B)$ relation is the Weber blockade of vortices \cite{Pekker2011-yq,Morgan-Wall2015-qf}. Under vortex-charge duality, the WL can be viewed as a vortex analog of a Coulomb-blockaded quantum dot. As field $B$ changes, the device periodically enters and exits the ``blockaded" states with fixed number of vortices, resulting in $I_c$ oscillations as a function of $B$ \cite{Pekker2011-yq}. $I_c(B)$ oscillations can also be seen in Devices A through C under certain measurement configurations (Figure~\ref{ABC IcB}(f), Figure~\ref{B config} and Figure~\ref{C config}). 

\section{Outlook}
The demonstration of a geometrically engineered, reconfigurable supercurrent diode effect (SDE) at the LAO/KTO interface establishes a uniquely versatile platform for both fundamental physics and quantum technologies. The ability to write, erase, and rewrite weak links with c-AFM lithography enables the rapid prototyping of non-reciprocal circuit elements for dissipationless electronics. This precise geometric control also enables on-demand engineering of energy landscape of vortices, making the system an ideal laboratory for systematic studies of 2D vortex dynamics. For example, a vortex pinning center may be defined by simply engaging a negatively-biased AFM tip on the device. Furthermore, large permittivity of KTO substrate enables electrostatic gating as a convenient tuning nob for devices. This, along with the large kinetic inductance of LAO/KTO interface \cite{Mallik2022-lx, Yang2025-kr}, results in slow light speed in the system which is crucial for compact circuit elements. This work positions the LAO/KTO system at the forefront of research into 2D SDE, vortex physics, and the next generation of quantum circuits. 

\section{Methods}
\subsection*{Growth of LAO on KTO (111) substrate}
The LaAlO$_3$ growth on KTaO$_3$ (111) substrate is carried out by pulsed laser deposition (PLD) with substrate heater temperature at 673 K in a dynamic oxygen pressure of 10$^{-5}$ torr. The laser has fluence of 1.6 J/cm$^2$ and repetition rate of 1 Hz (248 nm, LPX 300, Coherent). LaAlO$_3$ is deposited from a single-crystal LaAlO$_3$ target (Crystec) with a target-to-substrate distance of 65 mm. The growth rate of LaAlO$_3$ is approximately 0.11 Å per laser pulse. Following the growth of 4.4 nm of LAO, the samples are cooled to room temperature by quenching in the growth atmosphere.
\subsection*{Conductive Atomic Force Microscope lithography}
We closely follow the c-AFM lithographic process in Ref.\cite{Yu2025-ey} to create WLs, except for the specific tip voltage $V_{tip}$. Here, the 2D channel and the leads are written by $V_{tip}\in[+20\  \mathrm{V},+30\ \mathrm{V}]$. Then the 2D channel is cut with $V_{tip}\in[-9\  \mathrm{V},-8\ \mathrm{V}]$ for 3-4 times until no conductance is left between the two halves. Finally the WL is written once with $V_{tip}\in[+7\  \mathrm{V},+8\ \mathrm{V}]$.
\subsection*{Current-voltage characteristics}
Low-temperature $I-V$ characteristics are measured in a Quantum Design Physical Property Measurement System (PPMS) with a dilution refrigerator (DR) unit. In PPMS, $B$ field perpendicular to the sample plane can be applied. Source voltages are output by National Instruments PXI-4461, which can perform both digital-to-analog and analog-to-digital conversion. Current biasing is achieved by shunting the device with 300 k$\Omega$ in-series resistance. The drain current and the voltages are measured after amplification by a Krohn-Hite 7008 multichannel preamplifier. When taking a single $I-V$ curve, the bias current $I$ ramps from 0 to the positive maximum, then to the negative minimum, and finally back to 0. This way both the switching current $I_c$ and the retrapping current $I_r$ are captured. No averaging is performed between the $I-V$ curves, and each datapoint in the $I_c(B)$, $I_r(B)$ and $\eta(B)$ plots is extracted from a single $I-V$ curve.
\subsection*{Time-dependent Ginzburg Landau simulations}
TDGL simulation is performed with the pyTDGL package (https://py-tdgl.readthedocs.io, Ref.\cite{Bishop-Van_Horn2023-vs}), which we modified to incorporate thermal fluctuations. In the pyTDGL package we first choose the following parameters for LAO/KTO(111) interface: Ginzburg-Landau correlation length $\xi_{GL}=20\,\nm$, normal state conductivity $\sigma=0.3\,\mathrm{S/\mu m}$, London penetration depth $\lambda_{L}=2.5\,\mathrm{\mu m}$, thickness of the LAO/KTO 2d electron gas (2DEG) $d=5$~nm and reduced temperature $t=T/T_c=0.05$. We justify these choices in the following:\\
(1) $\xi_{GL}$ is extracted from the out-of-plane critical field $B_{c2}(T=0)=\Phi_0/(2\pi\xi_{GL}^2)$ of a Hall-bar device in Figure~\ref{2d}. At $V_{bg}=0$~V, $\xi_{GL}=21.2\,\nm$ while at $V_{bg}=-60$~V, $\xi_{GL}$ decreases to 16.2~nm. In the TDGL simulation we define $\xi_{GL}=20\,\nm$ which is close to the measured value at $V_{bg}=0$~V.\\
(2) The thickness $d$ of the LAO/KTO(111) 2DEG has been calibrated in Ref.\cite{Liu2021-eq} to be 5.1~nm, and also reported by Ref.\cite{Chen2021-dj} to vary from 2~nm to 6~nm depending on the $V_{bg}$ applied. Here we choose $d=5\,\nm$ in our simulation.\\
(3) Normal state conductivity can also be extracted from Figure~\ref{2d}. The Hall-bar with 4:1 ratio has normal state resistance $R_N=2.6\ \mathrm{k\Omega}$ at $V_{bg}=0$~V and $R_N=5\ \mathrm{k\Omega}$ at $V_{bg}=-60$~V. We choose the $R_N$ at $V_{bg}=0$~V, which gives sheet resistance $R_{sheet}=650\ \Omega$, resistivity $\rho=R_{sheet}d=3.25\ \Omega\cdot\mu$m and conductivity $\sigma=1/\rho\approx0.3$ S/$\mu$m.\\
(4) $\lambda_L$ is related to the 3d superfluid density $n_{s,3d}$ in the following way: $\lambda_{L}^2=m/(\mu_0n_{s,3d}e^2)$. The 2d superfluid density of LAO/KTO(111) has been reported in Ref.\cite{Mallik2022-lx} to be $n_{s,2d}=2\times10^{12}$~cm$^{-2}$. Thus we can estimate $\lambda_L=\sqrt{md/(\mu_0n_{s,2d}e^2)}\approx2.5\ \mu$m.\\
(5) Thermal noise terms that depend on $t=T/T_c$ are included in the TDGL equations during simulation. $T_c$ of the LAO/KTO(111) interface is known to be 1 to 2 K (Ref.~\cite{Liu2021-eq} and Figure~\ref{D temp}), and the $I-V$ measurements are performed in a PPMS setup with base temperature $T=50$~mK, so $t$ is chosen to be 0.05. We note that the actual electron temperature in our devices may be slightly higher than the cryostat temperature, which needs future noise thermometry to be accurately measured.\\

The pyTDGL package then generates the finite volume mesh for the device in Figure~\ref{tdgl}(a). We specify the maximum edge length to be 14~nm which is smaller than $\xi_{GL}$, resulting in $\approx6000$ mesh points. PyTDGL simulates how the order parameter $\psi$ evolves at each mesh point as a function of time. It outputs the phase difference and voltage across the two probe points we define (Figure~\ref{tdgl}(d)). The time unit $\tau_0$ of the horizontal axis in Figure~\ref{tdgl}(d) has the value $\tau_0=\mu_0\sigma\lambda_{L}^2=2.4$~ps. and the voltage unit $V_0$ of the vertical axis has the value $V_0=2\Phi_0/(\pi\tau_0)=0.56$~mV. At $B=-2000\,\Oe$ and $I=\pm150\,\nA$, the pyTDGL solver first goes through a ``thermalization" step which lasts for $T_{therm}=550\tau_0$, where the device is stabilized at the set field and current bias. Then the solver solves for a duration $T_{solve}=650\tau_0$ while recording phase $\Delta\phi(t)$ and voltage $V(t)$ to be plotted in Figure~\ref{tdgl}(d). The current density plots Figure~\ref{tdgl}(b)(c) is recorded at the timestamp $t=100\tau_0$ during the solving step.\\

We run the pyTDGL solver at a series of current values to get a simulated $I-V$ curve: $I$ from 0~nA to -210~nA with step of -1.5~nA, and then from 0~nA to +210~nA with step of 1.5~nA. With this $I$ sequence we capture the switching current at both positive and negative bias to simulate the SDE strength correctly. At each current value, the solver thermalizes for $T_{therm}=80\tau_0$ and then solve for $T_{solve}=90\tau_0$. The mean voltage within this $90\tau_0$ solving time is recorded as the DC voltage to be plotted in Figure~\ref{tdgl}(e)(f). The solution of the previous $I$ is used as seed solution for the next $I$ for faster thermalization. This $I-V$ curve simulation is then repeated at a series of $B$ from $-6000\,\Oe$ to $6000\,\Oe$ with a step of $250\,\Oe$ to get the $V$ vs $I$ vs $B$ plot (Figure~\ref{tdgl}(e)). We note the simulated $I_c(B=0)=190\,\nA$ (Figure~\ref{tdgl}(g)) agrees well with the experimentally measured $I_c(B=0)$ from Devices A through F, which ranges from 120~nA to 210~nA.

\begin{acknowledgement}
This work was supported by the National Science Foundation (NSF) Grant No. DMR-2225888 (J.L. and P.I.) and by the Defense Advanced Research Projects Agency (DARPA) under Agreement No. HR00112490317 (D.P. and J.L.). Thin film synthesis at the University of Wisconsin–Madison was supported by the US Department of Energy (DOE), Office of Science, Office of Basic Energy Sciences (BES), under award number DE-FG02-06ER46327. C.B.E. acknowledges support for this research through a Vannevar Bush Faculty Fellowship (ONR N00014-20-1-2844), and the Gordon and Betty Moore Foundation’s EPiQS Initiative, Grant GBMF9065.  
\end{acknowledgement}

\begin{suppinfo}
Supplementary Information (PDF) including Supplementary Figures S1-S16 and Supplementary Notes 1-3. The complete supplementary material is available from the original manuscript.
\end{suppinfo}

\bibliography{text}


\setcounter{secnumdepth}{3}  
\renewcommand{\thesection}{S\arabic{section}}
\renewcommand{\thefigure}{S\arabic{figure}}
\renewcommand{\thetable}{S\arabic{table}}
\renewcommand{\theequation}{S\arabic{equation}}
\renewcommand{\thepage}{S\arabic{page}}
\setcounter{figure}{0}
\setcounter{table}{0}
\setcounter{equation}{0}
\setcounter{page}{1}
\setcounter{section}{0}

\clearpage
\begin{center}
\Large\textbf{Supporting Information for:}\\[0.5em]
\Large\textbf{KTaO$_3$-Based Supercurrent Diode}

\vspace{1em}

\normalsize
Muqing Yu\(^1\),
Jieun Kim\(^2\),
Ahmed Omran\(^1\),
Zhuan Li\(^1\),
Jiangfeng Yang\(^2\),
Sayanwita Biswas\(^1\),
Chang-Beom Eom\(^2\),
David Pekker\(^1\),
Patrick Irvin\(^1\),
Jeremy Levy\(^{1,*}\)

\vspace{0.5em}

\small
\(^1\)Department of Physics and Astronomy, University of Pittsburgh, Pittsburgh, PA 15260, USA\\
\(^2\)Department of Materials Science and Engineering, University of Wisconsin-Madison, Madison, WI 53706, USA\\
\(^*\)Corresponding author. Email: jlevy@pitt.edu

\vspace{1.5em}

\textbf{This PDF file includes:}\\
Supplementary Notes 1-3\\
Figures S1 to S16
\end{center}

\vspace{2em}

\section{Supplementary Note S1: Effect of electrostatic gating on the SDE of Devices A-C}
\label{SI:gate}

KTO is known to be quantum paraelectric at cryogenic temperature~\cite{Fujii1976-ph}, which enables tuning of superfluid density and disorder in its two-dimensional electron gas by applying a voltage $V_{\rm bg}$ on the backside of the sample~\cite{Chen2021-dj}. We note the plots in Figure~2 of the main text are taken with $V_{\rm bg}=-30$~V applied on Devices A-C. For each device, $I_{c\pm}(B)$ and $\eta(B)$ are also measured with $V_{\rm bg}=-55$~V and with $V_{\rm bg}=0$~V, shown in Figure~\ref{ABC IcB -50} and Figure~\ref{ABC IcB 0} respectively. From the intensity plots of $dV/dI$ versus $I$ versus $B$ (Figure~\ref{ABC IcB 0}(a)-(c), main text Figure~2(a)-(c), Figure~\ref{ABC IcB -50}(a)-(c)), we clearly observe the increase in $dV/dI$ in the normal state when $V_{\rm bg}$ decreases from 0~V to $-30$~V to $-55$~V.

In terms of the diode behavior of each device, applying different $V_{\rm bg}$ does not change its SDE polarity. Device~B always exhibits $\eta<0$ at $B>0$ and $\eta>0$ at $B<0$ (Figure~\ref{ABC IcB 0}(h), main text Figure~2(h), Figure~\ref{ABC IcB -50}(h)). Device~C always shows the opposite polarity compared to Device~B, with $\eta>0$ at $B>0$ and $\eta<0$ at $B<0$ (Figure~\ref{ABC IcB 0}(i), main text Figure~2(i), Figure~\ref{ABC IcB -50}(i)). The reference Device~A always exhibits weaker SDE with $|\eta|<4\%$ (Figure~\ref{ABC IcB 0}(g), main text Figure~2(g), Figure~\ref{ABC IcB -50}(g)).

However, $V_{\rm bg}$ can affect the specific $I_{c\pm}(B)$ and $\eta(B)$ patterns as well as extreme values of $\eta$. Figure~\ref{ABC bg} shows how $\eta_{\rm max}$ and $\eta_{\rm min}$ as well as the corresponding optimal $B$ field $B_{\eta \rm max}$ and $B_{\eta \rm min}$ evolve with $V_{\rm bg}$. Highest $\eta_{\rm max}$ and $|\eta_{\rm min}|$ of Device~C are achieved at $V_{\rm bg}=0$~V, while $\eta_{\rm max}$ and $|\eta_{\rm min}|$ of Device~B increase at negative $V_{\rm bg}$. The parameters $B_{\eta \rm max}$ and $B_{\eta \rm min}$ of Device~B change non-monotonically with $V_{\rm bg}$. This inconsistency between Devices B and C prevents us from reaching any solid conclusion on how $V_{\rm bg}$ affects the diode performance.

\section{Supplementary Note S2: Estimation of dimensions of KTO WLs}
\label{SI:dimensions}

Devices A-E are patterned by cutting the two-dimensional conducting channel in two halves and then bridging them back together by writing a~nanowire. In this section, we provide an estimation of length $l_{\rm WL}$ and width $w_{\rm WL}$ for the resulting WLs. Ref.~\citenum{Yu2025-ey} extracted the current-phase relationship (CPR) of KTO WLs by measuring quantum interference between two parallel WLs. From the CPR, $l_{\rm WL}$ is determined to be 200 to 300~nm, varying from Device~to device. The c-AFM lithographic process in this work closely follows Ref.~\citenum{Yu2025-ey}, so we believe $l_{\rm WL}\in[200\,\nm,300\,\nm]$ also applies to the WLs we create here. For the TDGL simulation described in main text Figure~4(a), we choose the lower bound $l_{\rm WL}=200\,\nm$.

In terms of $w_{\rm WL}$, we choose it to be 50~nm in the TDGL simulation, which can be justified by Device~F shown in Figure~\ref{F}. In Device~F, instead of cutting the 2D channel completely in half, we leave a 60 nm gap at the bottom (Figure~\ref{F}(a)), which is effectively a WL with $w_{\rm WL,F}\approx60\,\nm$. At $T=50$~mK, $B=0$ with $V_{\rm bg}=-30$~V applied, its critical currents $I_c=I_{c+}=|I_{c-}|=205\,\nA$ (Figure~\ref{F}(d)), as there is no SDE at $B=0$. We can compare this value to Devices A-C: (1) Device~A, $I_c(B=0)=212\,\nA$; (2) Device~B, $I_c(B=0)=198\,\nA$; (3) Device~C, $I_c(B=0)=220\,\nA$, which are all measured at $T=50$~mK with $V_{\rm bg}=-30$~V applied (main text Figure~2(d)(e)(f)). For Device~D and E: (4) Device~D, $I_c(B=0)=132\,\nA$ (Figure~\ref{D bg}(b)); (5) Device~E, $I_c(B=0)=120\,\nA$ (Figure~\ref{E bg}(b)), both of which are measured at $T=50$~mK with $V_{\rm bg}=-40$~V.

Under similar measurement conditions, the averaged $I_c$ of WLs A-E is $\langle I_c(B=0)\rangle_{A-E}=175$~nA, which is $85\%$ of the $I_c$ of Device~F. If we~nAïvely assume that the measured $I_c$ is directly proportional to $w_{\rm WL}$, then the averaged width of WLs A-E $\langle w_{\rm WL}\rangle_{A-E}\approx85\%\times w_{\rm WL,F}\approx51$~nm. We note the above argument may not hold in a 2D superconducting system with $w_{\rm WL}\approx50$~nm$>\xi_{\rm GL}\approx20$~nm (see Figure~\ref{2d} for calibration of $\xi_{\rm GL}$), where dissipation is governed by the entrance/nucleation of vortices. Nonetheless, $I_c$ is still a monotonic function of $w_{\rm WL}$ in 2D. Since $\langle I_c\rangle_{A-E}=85\%\ I_{c,F}$, we can still argue that $\langle w_{\rm WL}\rangle_{A-E}$ is a bit less than $w_{\rm WL,F}\approx60\,\nm$. Thus, 50~nm is a credible expectation for the width of a typical WL created by the ``cutting + bridging'' lithographic process.

Moreover, by using $w_{\rm WL}=50$~nm in the TDGL simulation, the calculated $I_c$ arrives at $I_c(B=0)=190$~nA (main text Figure~4(g)), very close to the $\langle I_c(B=0)\rangle_{A-E}=175$~nA. In conclusion, we believe $l_{\rm WL}=200$~nm and $w_{\rm WL}=50$~nm to be a credible estimation for our WL dimensions.

\section{Supplementary Note S3: Dependence of $I_c(B)$ and $\eta(B)$ on measurement configurations}
\label{SI:config}

Each of Devices A-C has 6 leads that can be used as either current leads or voltage leads during $I$-$V$ measurements. Eight different measurement configurations have been used to probe Devices A-C, which are listed in Figure~\ref{config}. $I_{c\pm}(B)$ of Device~A changes subtly when measured by different configurations (Figure~\ref{A config}), with its $\eta$ always lying within $\pm4\%$. Meanwhile, $I_{c\pm}(B)$ patterns of Devices B and C shows obvious change upon switching configurations (Figure~\ref{B config}, Figure~\ref{C config}), as does the corresponding $\eta(B)$. Despite the remarkable changes in the specific $\eta(B)$ pattern, SDE in Device~B does not switch sign, maintaining $\eta(B<0)>0$ and $\eta(B>0)<0$ under all configurations (Figure~\ref{B config}). The same argument holds for Device~C, which always has $\eta(B<0)<0$ and $\eta(B>0)>0$ (Figure~\ref{C config}).

TDGL simulation provides qualitative explanations for the dependence of $I_c(B)$ and $\eta(B)$ on the choice of current leads. We simulate the device shown in main text Figure~4(a) again in Figure~\ref{tdgl2}, the difference being that the current leads are relocated to the right edge in the configuration in Figure~\ref{tdgl2}(a), and to the left edge in the configuration in Figure~\ref{tdgl2}(b). Current density \textbf{\textit{K}} is simulated under $B=-2000$~oe and $I=+150$~nA (Figure~\ref{tdgl2}(c)(d)), same as the condition of main text Figure~4(b). Distribution of \textbf{\textit{K}} changes upon switching to different current leads, so does the number and location of static vortices in the 2d channel (Figure~\ref{tdgl2}(c) vs (d)). The right-sided current leads result in mobile vortex entry and dissipation (blue curves, Figure~\ref{tdgl2}(e)). Meanwhile the left-sided current leads create a current profile that is less effective at forcing vortex entry near the WL (a higher surface barrier), causing absence of dissipation (orange curves, Figure~\ref{tdgl2}(e)).

Simulated $I-V$ curves at $B=-2000\,\Oe$ give $I_{c+}=156\,\nA$ for right-sided current leads, a close $I_{c+}=157.5\,\nA$ for the top/bottom current leads in Figure~\ref{tdgl}(a), and an increased $I_{c+}=163.5\,\nA$ for left-sided leads (Figure~\ref{tdgl2}(f)). Under negative bias, these three configurations have the same $I_{c-}=-172.5\,\nA$ (Figure~\ref{tdgl2}(g)). Simulated diode efficiency $\eta=+5.0\%\,,+4.5\%\,,+2.7\%$ for the right-sided, top/bottom and left-sided current leads, which differ in magnitude but maintain the same sign, consistent with the experimental observation in Figure~\ref{B config} and Figure~\ref{C config}. In conclusion, according to the TDGL simulation, choosing different current leads can change $I_c$ due to the alteration of current density and vortex surface barriers. We also note the thermalization time $T_{therm}=80\tau_0$ used to simulate $I-V$ curves results in a slightly higher $I_c$ in Figure~\ref{tdgl2}(f) as compared to Figure~\ref{tdgl2}(e), where only one current value is simulated with prolonged thermalization step $T_{therm}=550\tau_0$ for the device to fully stabilize (see Methods Section). 

Another possible but less straightforward reason for the dependence of $I_c(B)$ on the lead configuration is the phase slips occuring within the current leads, which may induce premature phase slips in the WL through certain nonlocal interactions (such as heating or AC Josephson effect).

\begin{figure}[p]
\centering
\includegraphics[scale=1]{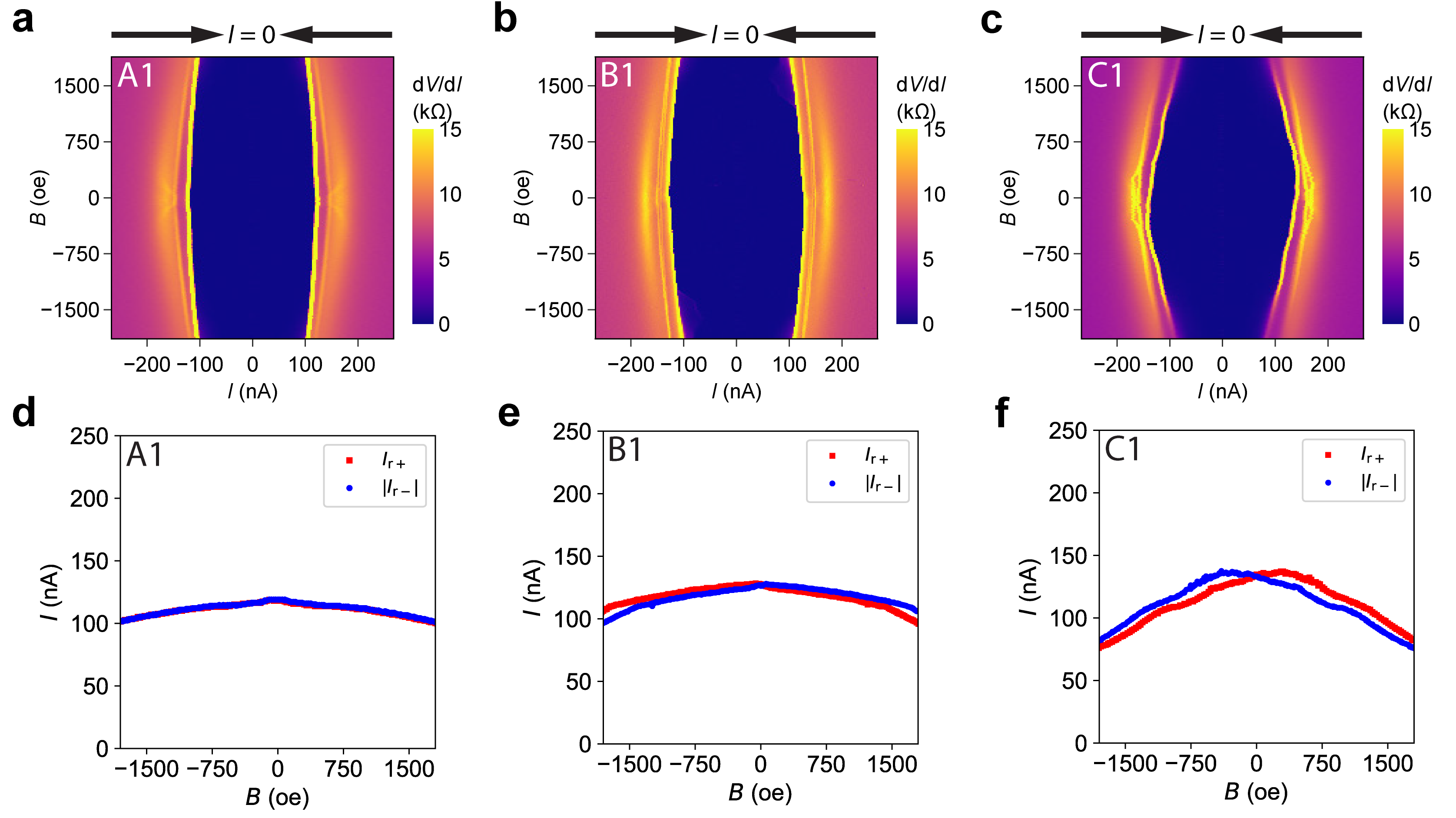}
\caption{Retrapping currents of Devices A-C vs magnetic field. (a)(b)(c) Intensity plots of differential resistance $dV/dI$ vs $I$ vs $B$ of Devices A, B and C. In these plots, current $I$ sweeps from $|I|>0$ to $I=0$, as indicated by the black arrows above each plot. (d)(e)(f) Extracted retrapping currents $I_{r\pm}$ of Devices A-C. These plots are from the same dataset as main text Figure~2, which was taken at $T=50$~mK with a backgate voltage $V_{\rm bg}=-30$~V applied on Devices A-C.}
\label{ABC IrB}
\end{figure}

\begin{figure}[p]
\centering
\includegraphics[scale=1]{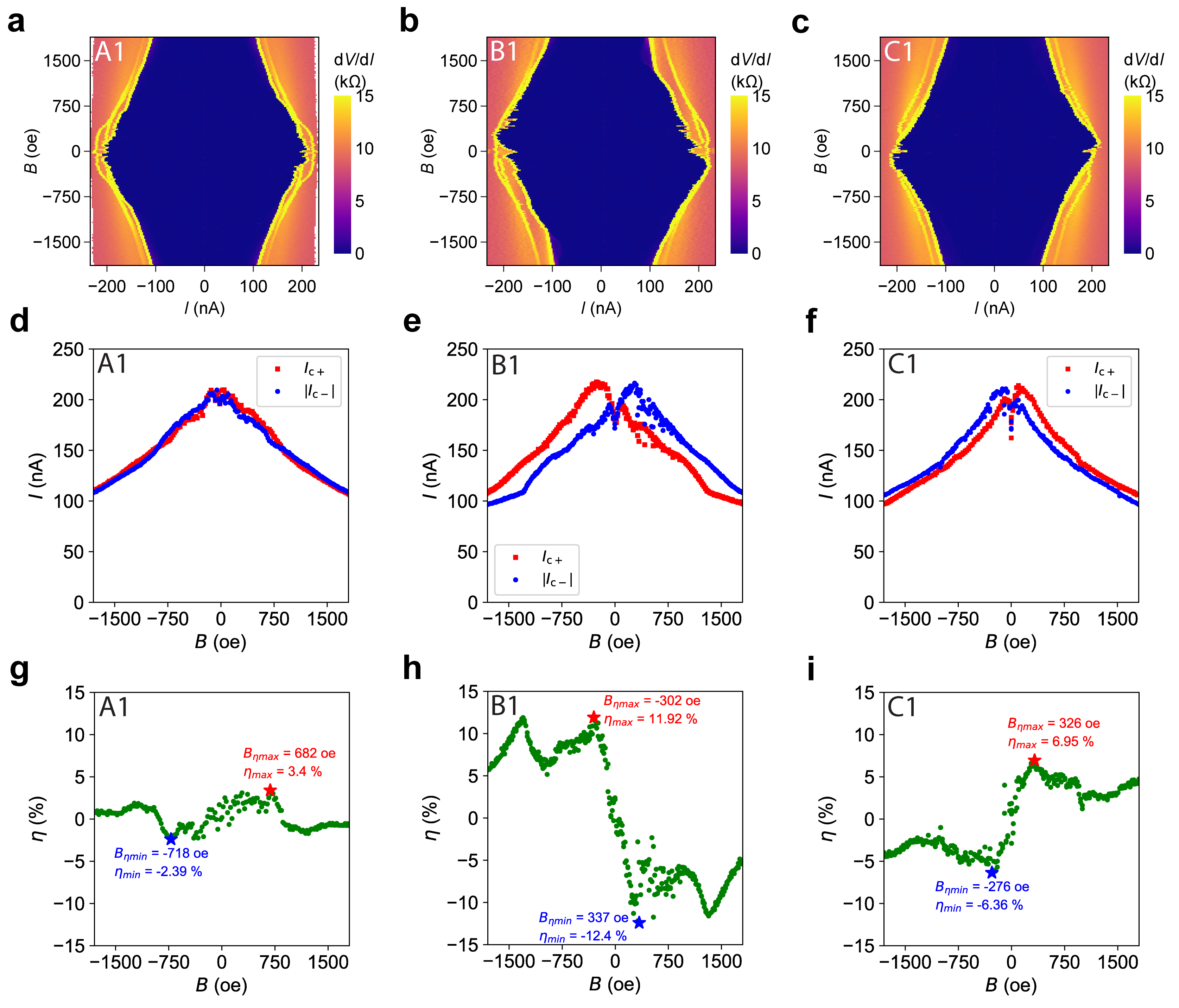}
\caption{Magnetic field sweep of Devices A-C taken at $V_{\rm bg}=-55$~V. (a)(b)(c) Intensity plots of $dV/dI$ vs $I$ vs $B$ of Devices A, B and C. (d)(e)(f) Switching current $I_{c\pm}$ of Devices A-C as a function of $B$. (g)(h)(i) Diode efficiency $\eta$ of Devices A-C as a function of $B$, with the locations of $\eta_{\rm max(min)}$ labeled. All plots in this figure were taken at $T=50$~mK with $V_{\rm bg}=-55$~V applied on Devices A-C.}
\label{ABC IcB -50}
\end{figure}

\begin{figure}[p]
\centering
\includegraphics[scale=1]{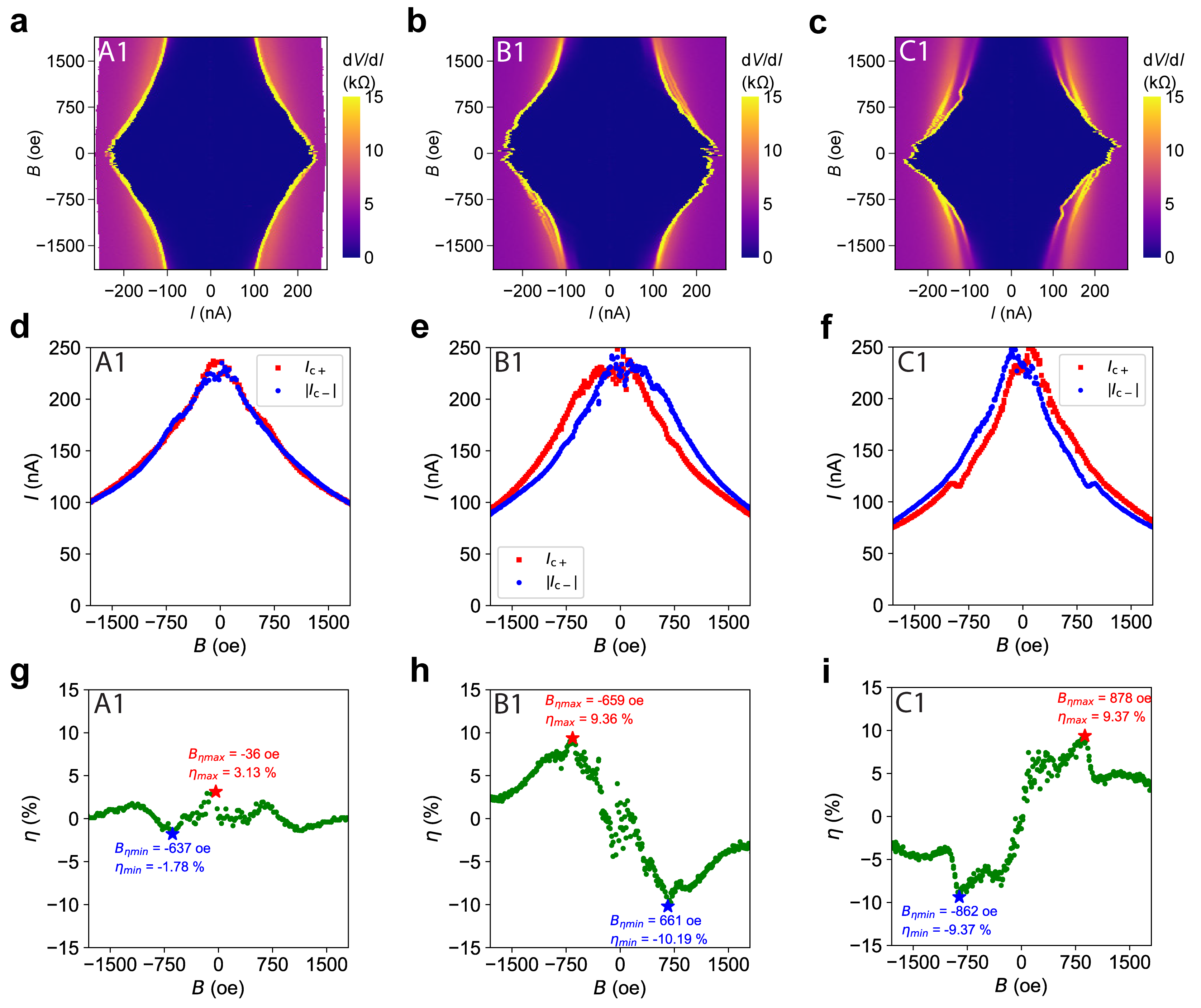}
\caption{Magnetic field sweep of Devices A-C taken at $V_{\rm bg}=0$~V. (a)(b)(c) Intensity plots of $dV/dI$ vs $I$ vs $B$ of Devices A, B and C. (d)(e)(f) Switching current $I_{c\pm}$ of Devices A-C as a function of $B$. (g)(h)(i) Diode efficiency $\eta$ of Devices A-C as a function of $B$, with the locations of $\eta_{\rm max(min)}$ labeled. All plots in this figure were taken at $T=50$~mK with backgate grounded ($V_{\rm bg}=0$~V).}
\label{ABC IcB 0}
\end{figure}

\begin{figure}[p]
\centering
\includegraphics[scale=0.6]{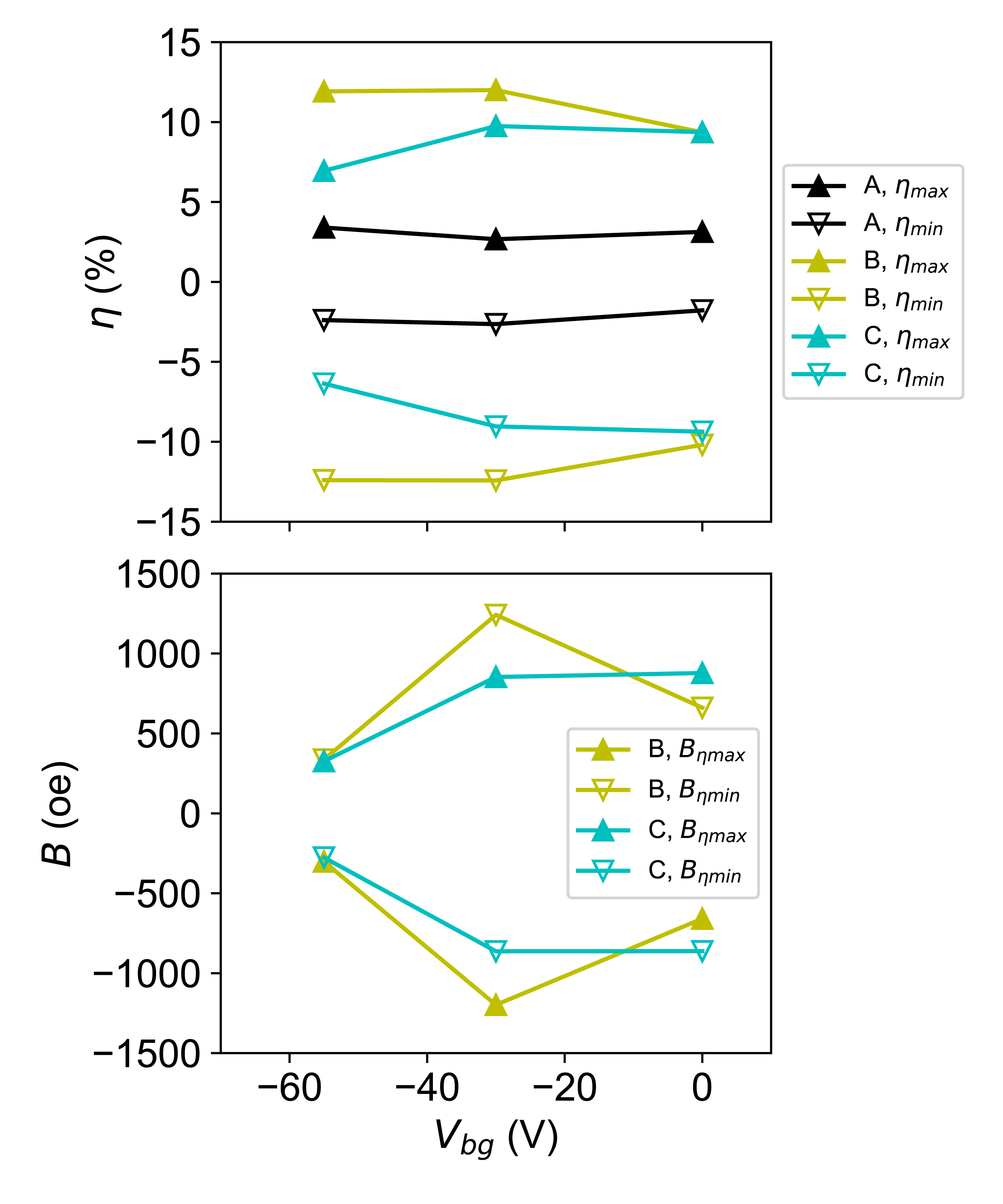}
\caption{Backgate dependence of diode efficiency of Devices A-C. Optimal diode efficiency $\eta_{\rm max(min)}$ (top) and corresponding optimal magnetic field $B_{\eta \rm min (\eta \rm max)}$ (bottom) are plotted as a function of $V_{\rm bg}$ applied on the sample. The datapoints are extracted from main text Figure~2(g)-(i) as well as Figure~\ref{ABC IcB -50}(g)-(i) and Figure~\ref{ABC IcB 0}(g)-(i).}
\label{ABC bg}
\end{figure}

\begin{figure}[p]
\centering
\includegraphics[scale=1]{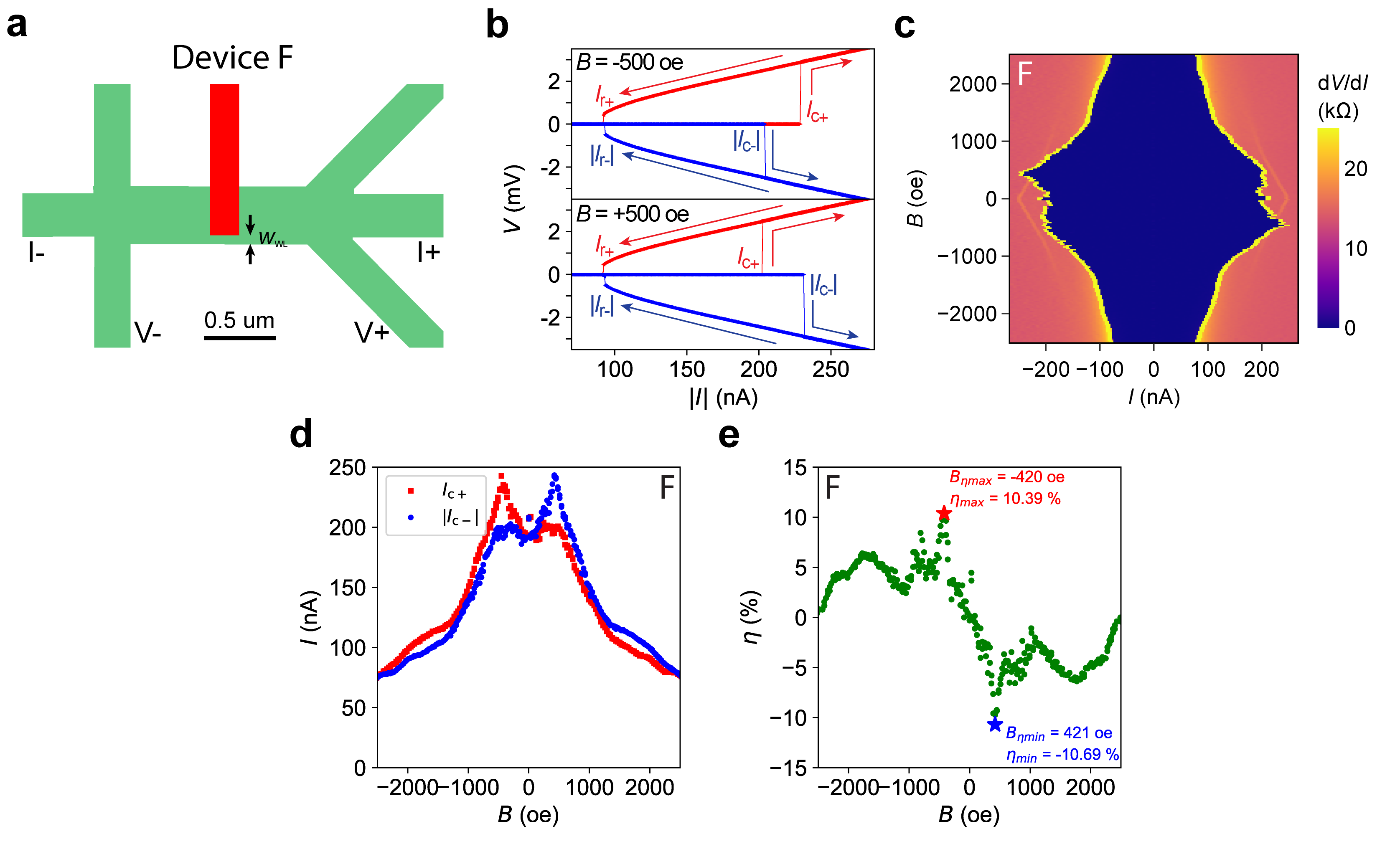}
\caption{Supercurrent Diode Device~F. (a) Layout of Device~F. Instead of cutting the 1D channel completely into the left and right halves and then writing the WL like Devices A-E, the 2D channel is only partially cut to leave a 60~nm-wide conducting path near the bottom edge. In this way we effectively create a WL with $w_{\rm WL}\approx60$~nm. (b) $I$-$V$ measurements of Device~F at $B=\pm500$~Oe, where arrows indicate the $I$ sweep direction. (c)(d)(e) $dV/dI$ vs $I$ vs $B$ intensity plot, $I_{c\pm}$ vs $B$ and $\eta$ vs $B$ plots of Device~F. All plots in this figure were taken at $T=50$~mK, with $V_{\rm bg}=-30$~V applied on the sample.}
\label{F}
\end{figure}

\begin{figure}[p]
\centering
\includegraphics[scale=1.25]{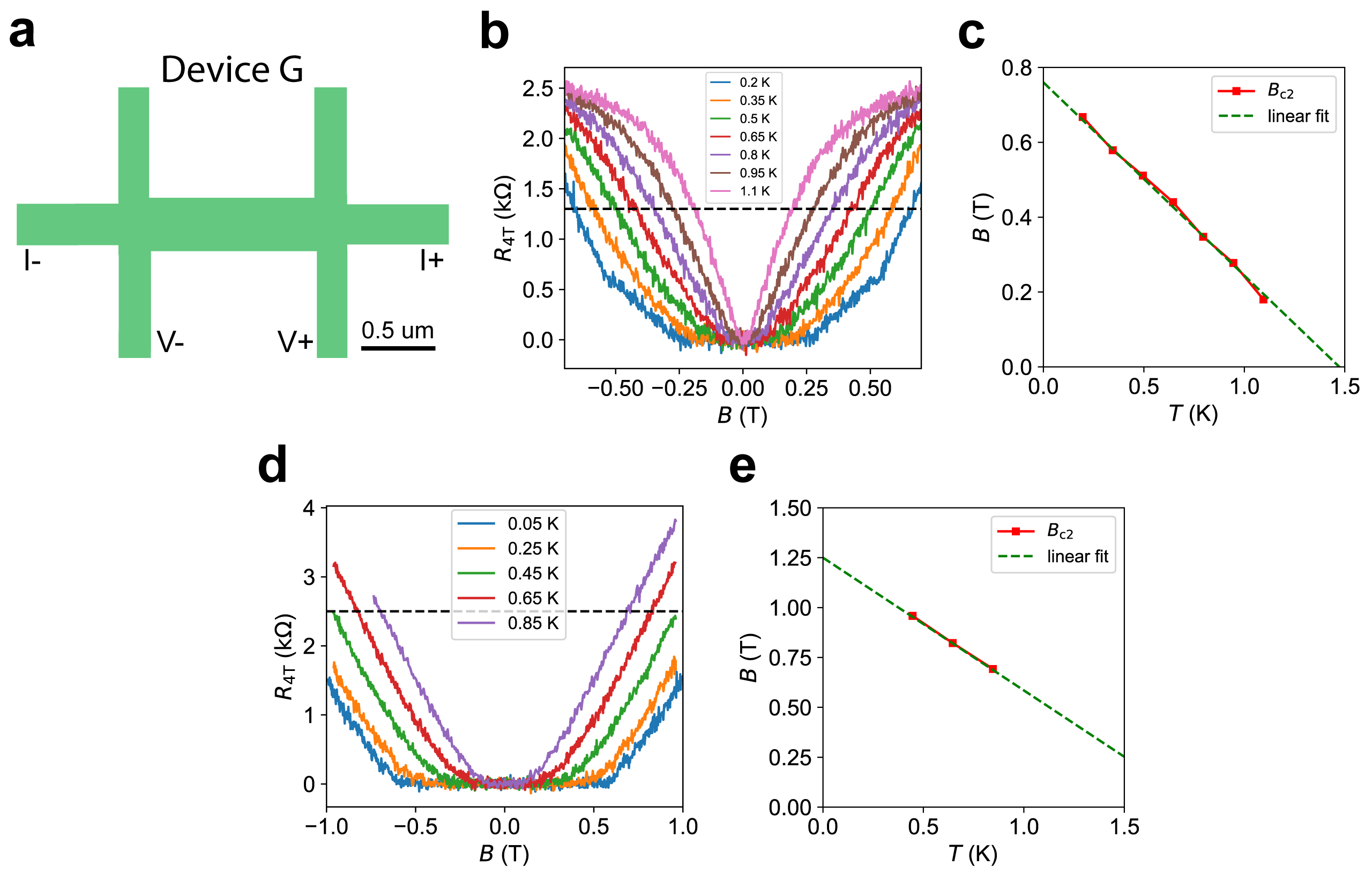}
\caption{Reference Hallbar Device~G. (a) Layout of Hallbar. (b) Four-terminal resistance $R_{\rm 4T}$ as a function of $B$ measured at different temperatures from 0.2 K to 1.1 K, with backgate grounded ($V_{\rm bg}=0$~V). The black dashed line indicates half of the normal state resistance $R_N/2$. (c) Upper critical field $B_{c2}$ as a function of $T$ at $V_{\rm bg}=0$~V. $B_{c2}$ at each temperature is extracted at $R_{\rm 4T}=R_N/2$ from panel (b). Performing linear fit using $B_{c2}(T)=\frac{\Phi_0}{2\pi\xi_{\rm GL}^2}(1-T/T_c)$ gives $B_{c2}(0)=0.73$~T and $\xi_{\rm GL}=21.2$~nm. (d) $R_{\rm 4T}$ as a function of $B$ measured at different temperatures from 0.05~K to 0.85~K, while applying $V_{\rm bg}=-60$~V on the sample. (e) Upper critical field $B_{c2}$ as a function of $T$ at $V_{\rm bg}=-60$~V, extracted from panel (d). Linear fit gives $B_{c2}(0)=1.25$~T and $\xi_{\rm GL}=16.2$~nm.}
\label{2d}
\end{figure}

\begin{figure}[p]
\centering
\includegraphics[scale=0.98]{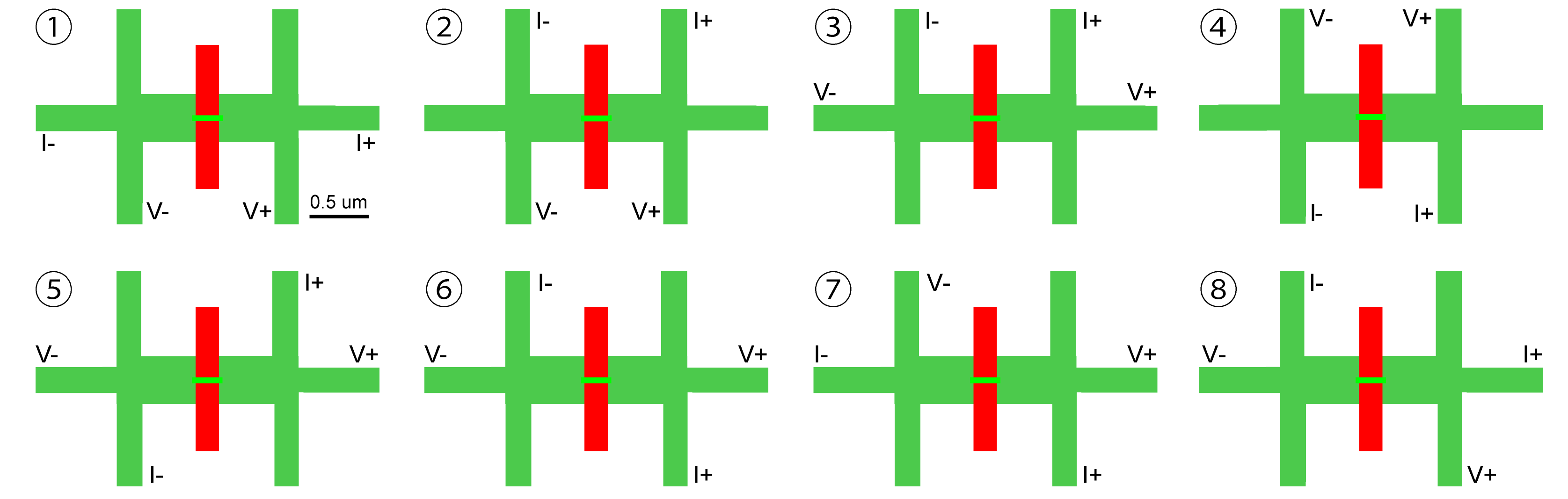}
\caption{Measurement configurations of Devices A-C. Each configuration is labeled with a number at its top right corner, which is referred to by $I$-$V$ measurements of Devices A-C. In each configuration, current source/drain are labeled by $I+$/$I-$, while the two voltage leads are labeled by $V+/V-$. We note that positive current $I>0$ always flow through the WL from right to left in all configurations.}
\label{config}
\end{figure}

\begin{figure}[p]
\centering
\includegraphics[scale=0.9]{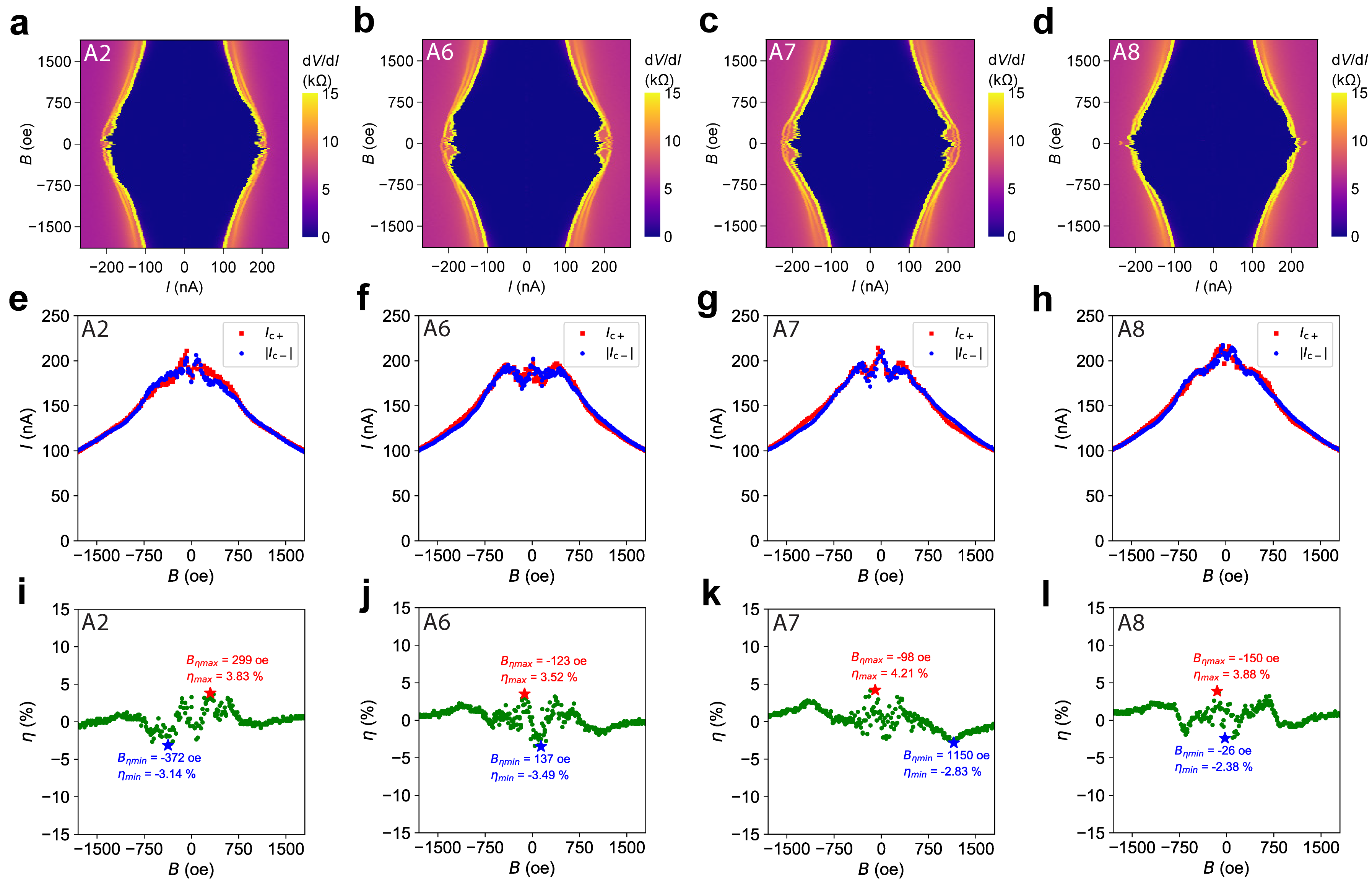}
\caption{Device~A under different measurement configurations. (a)-(d) $dV/dI$ vs $I$ vs $B$ plots of Device~A, with corresponding measurement configuration labeled at the top left corner of each plot (refer to Figure~\ref{config}). (e)-(h) $I_c$ vs $B$ of Device~A measured at different configurations. (i)-(l) Extracted $\eta$ vs $B$ of Device~A at different configurations. All plots in this figure were taken at $T=50$~mK with $V_{\rm bg}=-30$~V applied on Device~A.}
\label{A config}
\end{figure}

\begin{figure}[p]
\centering
\includegraphics[scale=0.85]{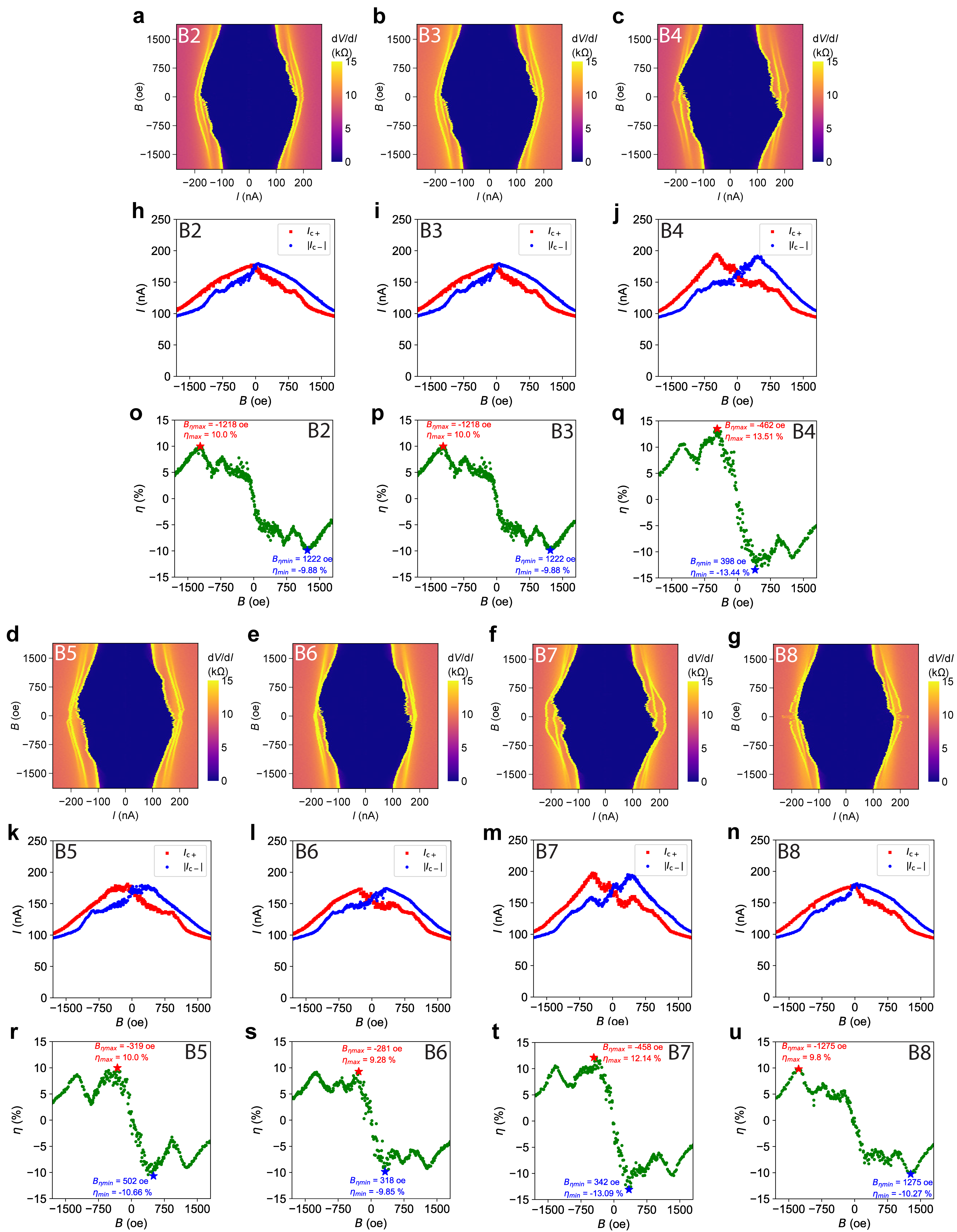}
\caption{Device~B under different measurement configurations. (a)-(g) $dV/dI$ vs $I$ vs $B$ plots of Device~B, with corresponding measurement configuration labeled at the top left corner of each plot (refer to Figure~\ref{config}). (h)-(n) $I_c$ vs $B$ of Device~B measured at different configurations. (o)-(u) Extracted $\eta$ vs $B$ of Device~B at different configurations. All plots in this figure were taken at $T=50$~mK with $V_{\rm bg}=-55$~V applied on Device~B.}
\label{B config}
\end{figure}

\begin{figure}[p]
\centering
\includegraphics[scale=0.85]{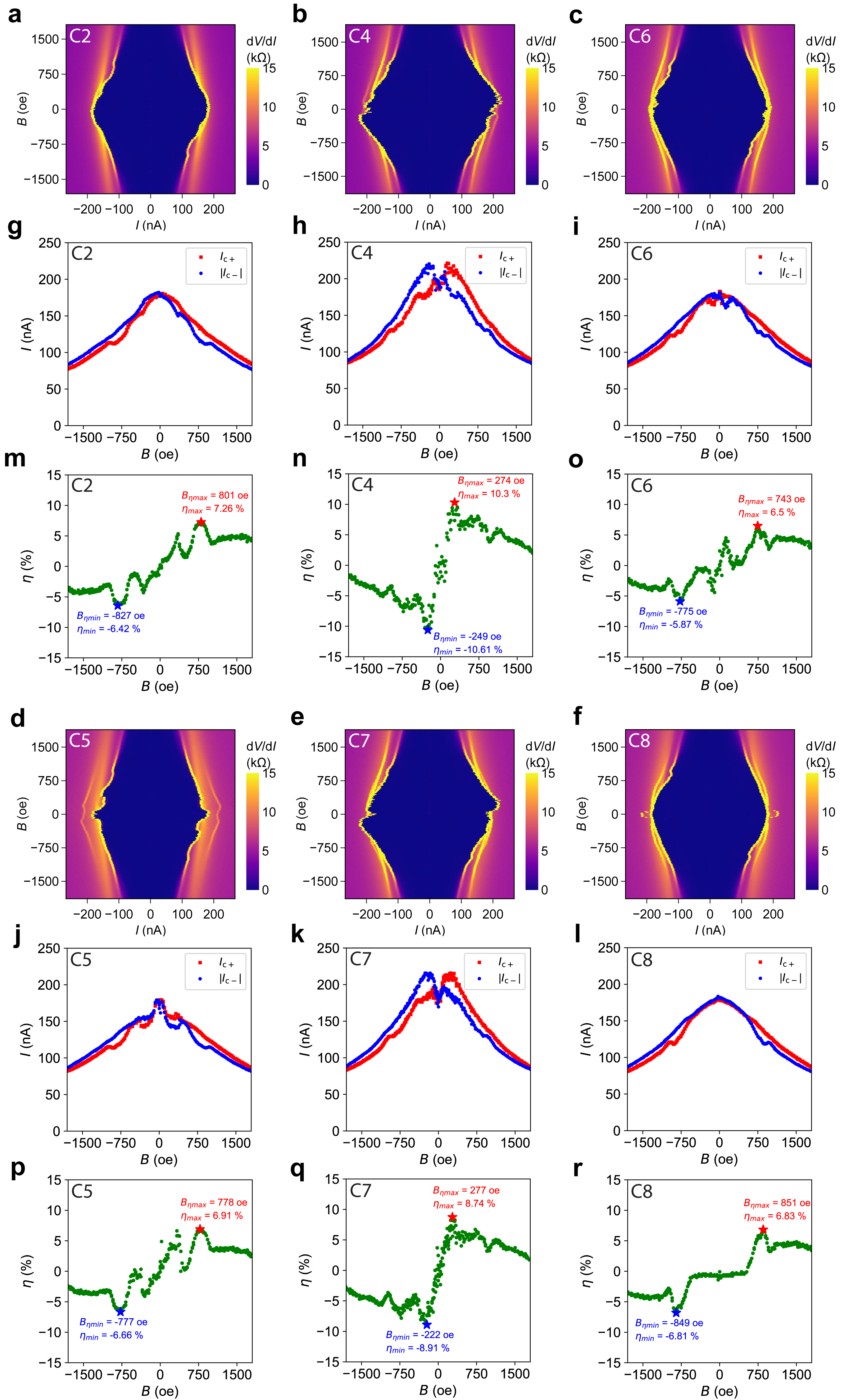}
\caption{Device~C under different measurement configurations. (a)-(f) $dV/dI$ vs $I$ vs $B$ plots of Device~C, with corresponding measurement configuration labeled at the top left corner of each plot (refer to Figure~\ref{config}). (g)-(l) $I_c$ vs $B$ of Device~C measured at different configurations. (m)-(r) Extracted $\eta$ vs $B$ of Device~C at different configurations. All plots in this figure were taken at $T=50$~mK with $V_{\rm bg}=-30$~V applied on Device~C.}
\label{C config}
\end{figure}


\begin{figure}[p]
\centering
\includegraphics[scale=1.1]{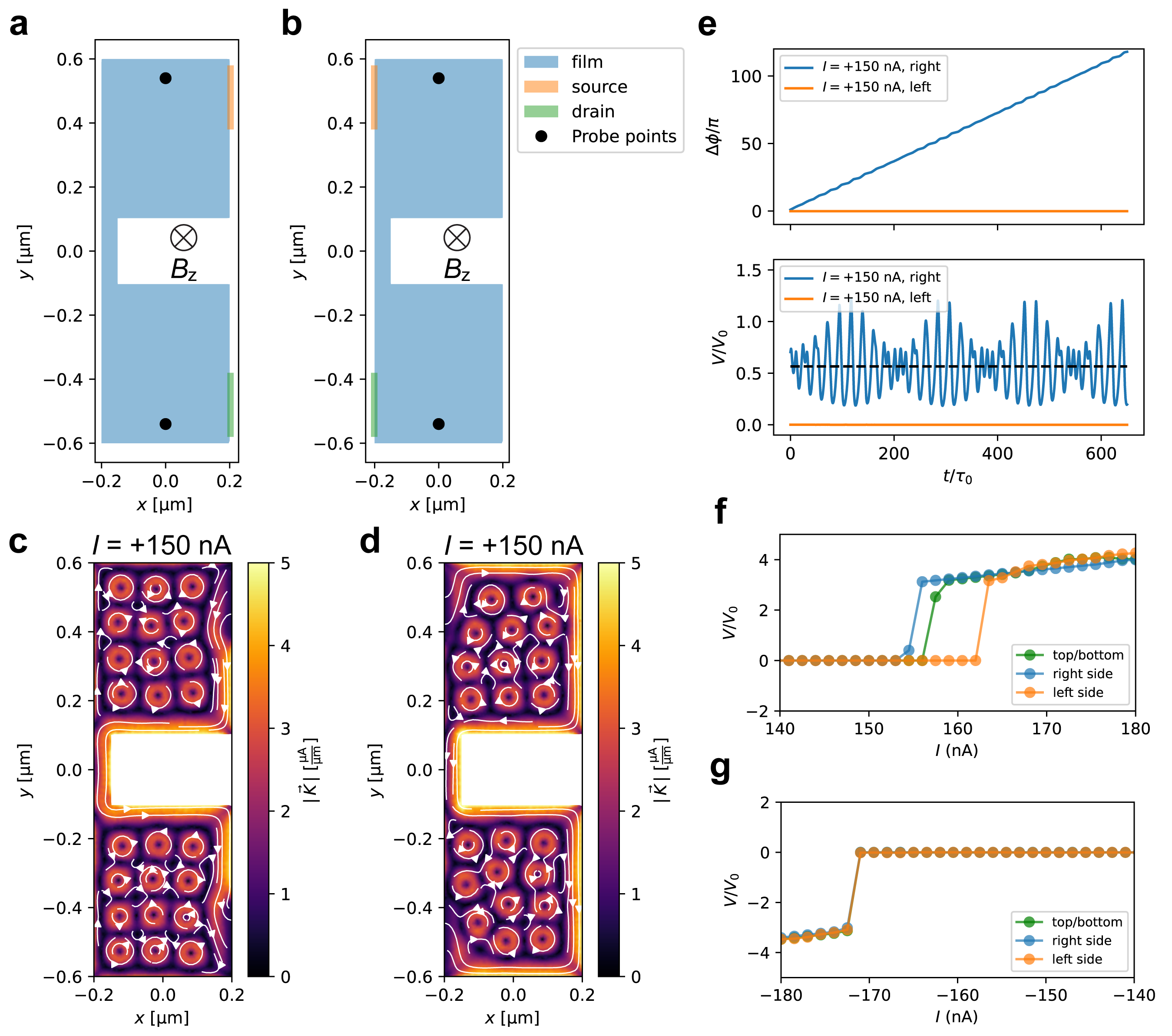}
\caption{TDGL simulation of the WL with different current sources and drains. (a) The simulated device is exactly the same as main text Figure~4(a), except here the current source and drain are put on the right edge of the device. (b) In this configuration the current source and drain are put on the left side. (c)(d) Current density \textbf{\textit{K}} simulated under $B=-2000\,\Oe$ field and $I=+150$~nA bias, using the configuration in panel (a) and (b) respectively. (e) Evolution of phase difference $\Delta \phi(t)$ and voltage $V(t)$. Black dashed line: time-averaged voltage of the configuration in (a). (f)(g) Simulated $I-V$ curves at $B=-2000\,\Oe$ using different configurations. Green curve is simulated using the configuration in main text Figure~4(a) where current source and drain are on the top and bottom edges. Blue and orange curves are simulated using the configurations in panel (a) (right-sided source and drain) and (b) (left-sided source and drain), respectively. }
\label{tdgl2}
\end{figure}


\begin{figure}[p]
\centering
\includegraphics[scale=1]{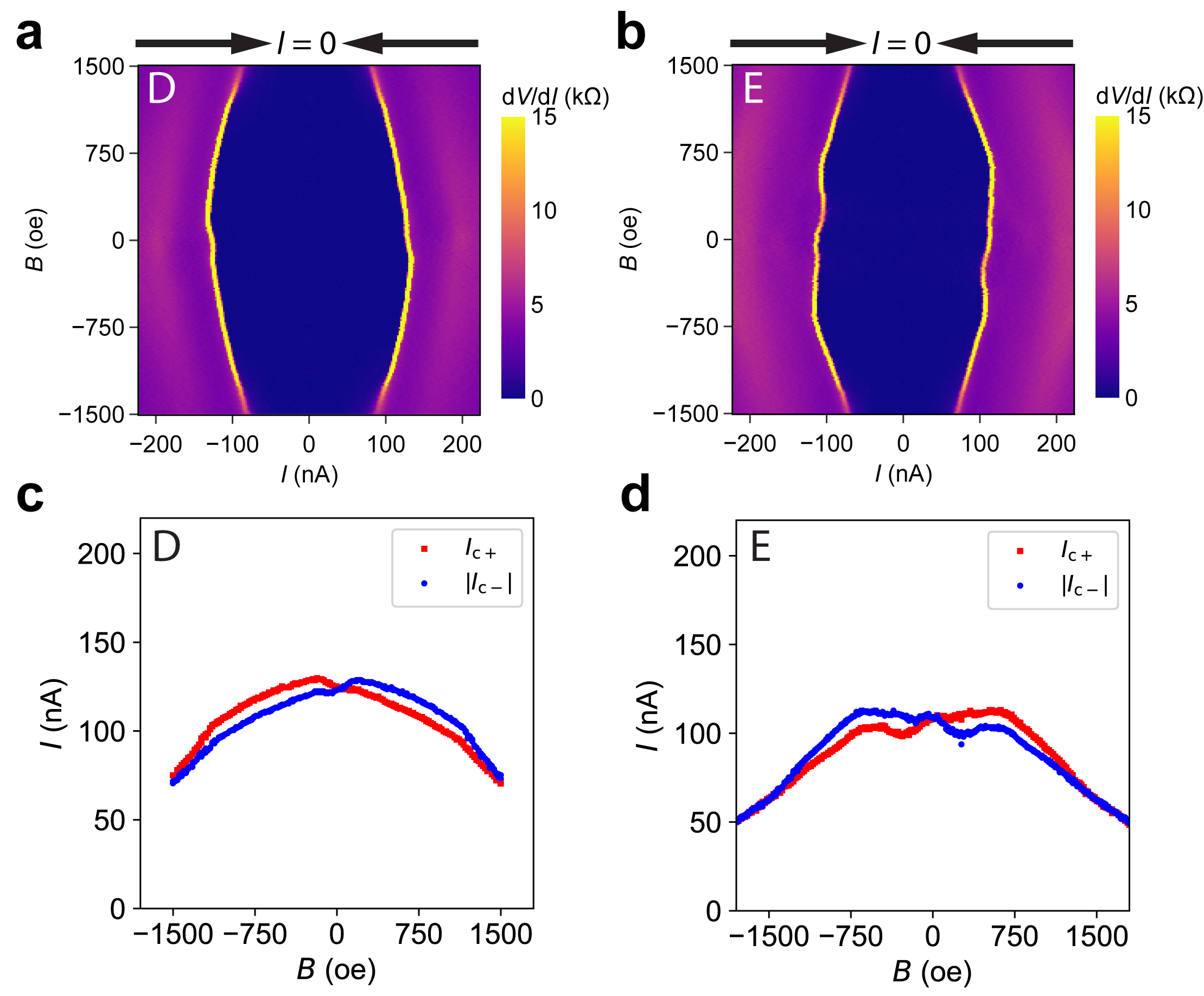}
\caption{Retrapping currents of Devices D and E vs magnetic field. (a)(b) Intensity plots of differential resistance $dV/dI$ vs $I$ vs $B$ of Devices D and E. In these plots, current $I$ sweeps from $|I|>0$ to $I=0$, as indicated by the black arrows above each plot. (c)(d) Extracted retrapping currents $I_{r\pm}$ of Devices D and E. These plots are from the same dataset as main text Figure~3, which was taken at $T=50$~mK with backgate grounded ($V_{\rm bg}=0$~V).}
\label{DE IrB}
\end{figure}

\begin{figure}[p]
\centering
\includegraphics[scale=0.97]{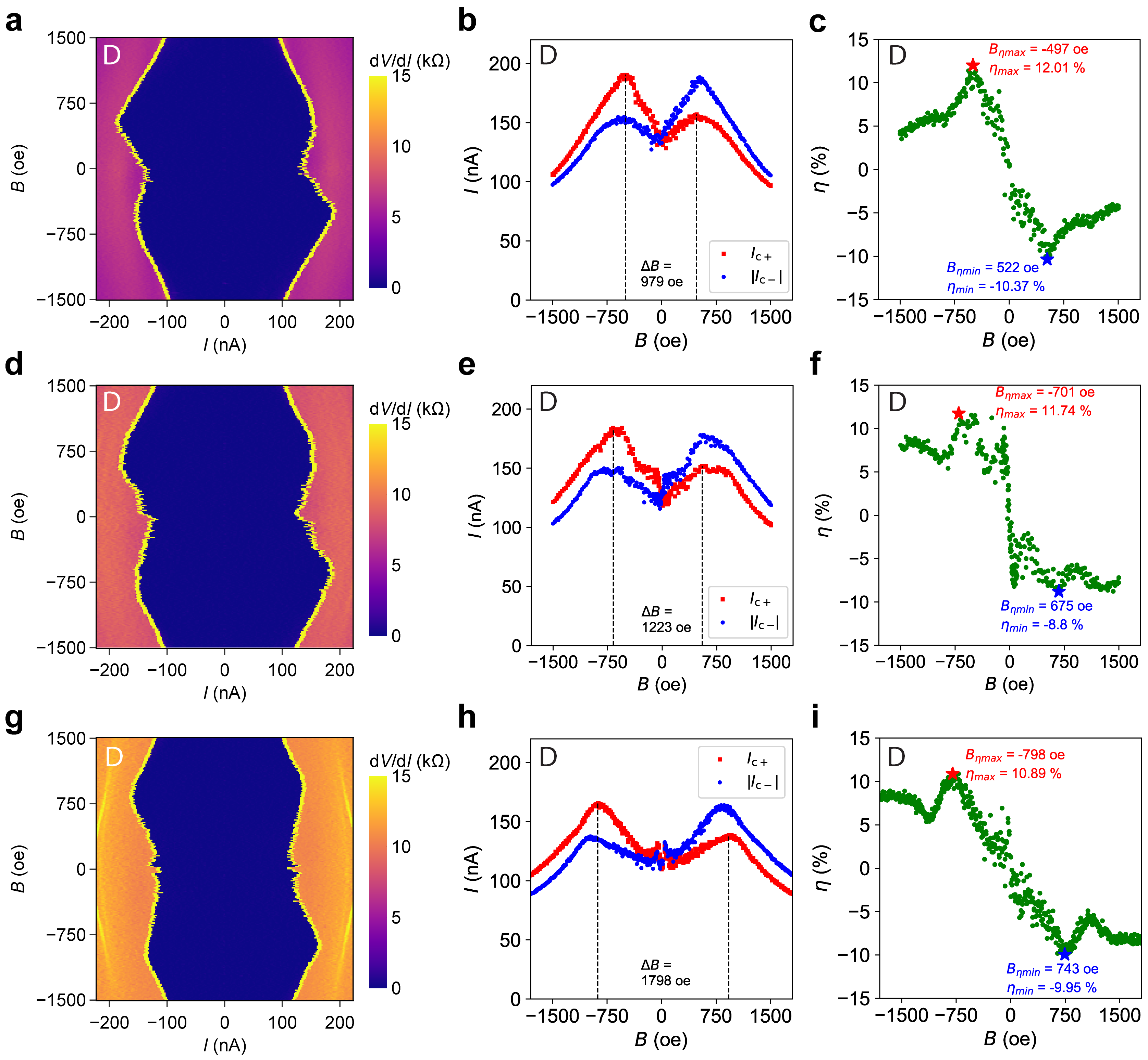}
\caption{Device~D measured at different backgate voltages. (a)(b)(c) $dV/dI$ vs $I$ vs $B$ intensity plot, $I_{c\pm}$ vs $B$ and $\eta$ vs $B$ relations of Device~D, measured at $V_{\rm bg}=-40$~V. The splitting $\Delta B$ between the two $I_c$ maxima is labeled, as well as the location of $\eta_{\rm max}$ and $\eta_{\rm min}$. (d)(e)(f) $dV/dI$ vs $I$ vs $B$ intensity plot, $I_{c\pm}$ vs $B$ and $\eta$ vs $B$ relations of Device~D, measured at $V_{\rm bg}=-60$~V. (g)(h)(i) $dV/dI$ vs $I$ vs $B$ intensity plot, $I_{c\pm}$ vs $B$ and $\eta$ vs $B$ relations of Device~D, measured at $V_{\rm bg}=-80$~V. All plots in this figure were taken at $T=50$~mK.}
\label{D bg}
\end{figure}

\begin{figure}[p]
\centering
\includegraphics[scale=0.97]{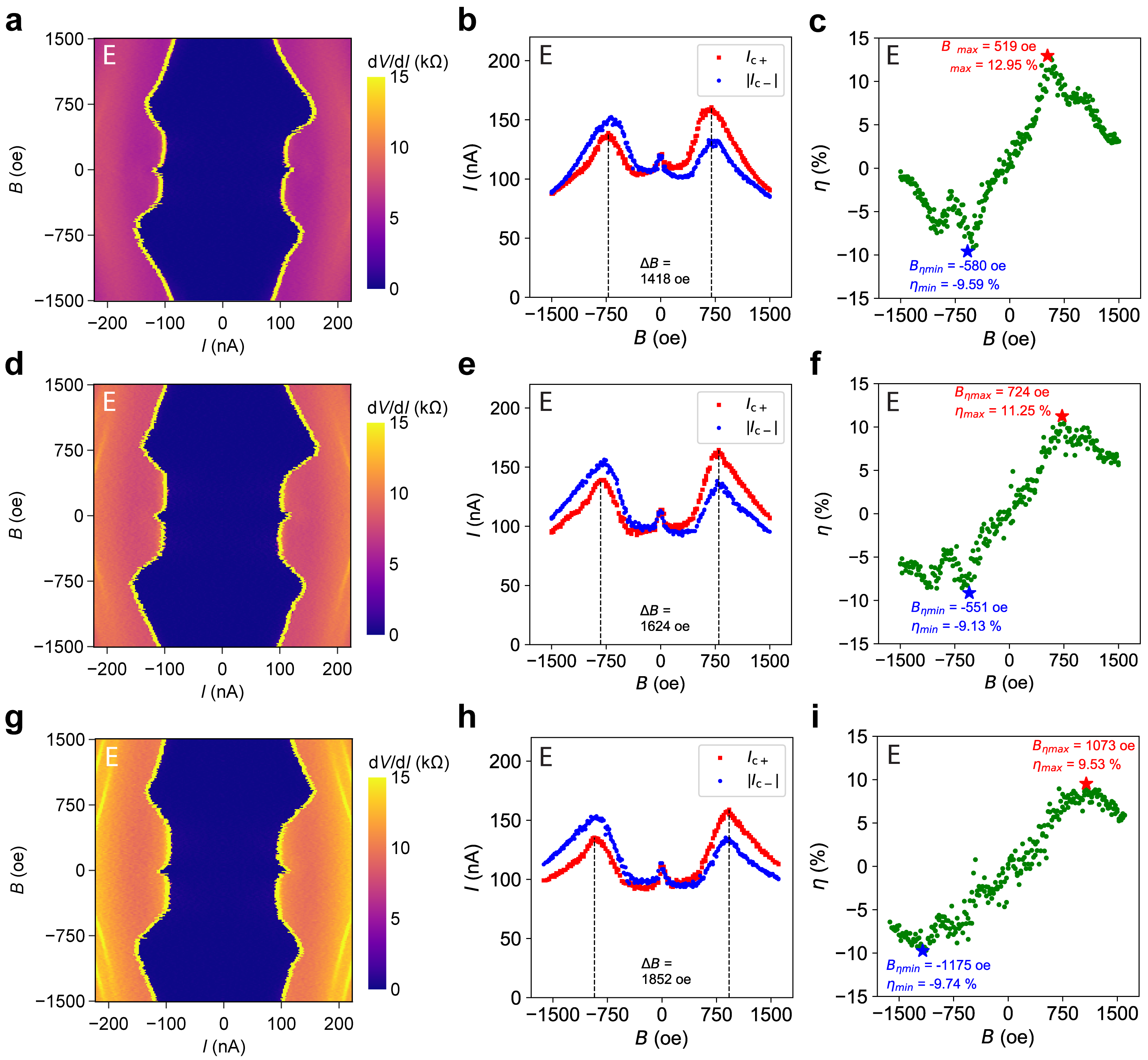}
\caption{Device~E measured at different backgate voltages. (a)(b)(c) $dV/dI$ vs $I$ vs $B$ intensity plot, $I_{c\pm}$ vs $B$ and $\eta$ vs $B$ relations of Device~E, measured at $V_{\rm bg}=-40$~V. The splitting $\Delta B$ between the two $I_c$ maxima is labeled, as well as the location of $\eta_{\rm max}$ and $\eta_{\rm min}$. (d)(e)(f) $dV/dI$ vs $I$ vs $B$ intensity plot, $I_{c\pm}$ vs $B$ and $\eta$ vs $B$ relations of Device~E, measured at $V_{\rm bg}=-60$~V. (g)(h)(i) $dV/dI$ vs $I$ vs $B$ intensity plot, $I_{c\pm}$ vs $B$ and $\eta$ vs $B$ relations of Device~E, measured at $V_{\rm bg}=-80$~V. All plots in this figure were taken at $T=50$~mK.}
\label{E bg}
\end{figure}

\begin{figure}[p]
\centering
\includegraphics[scale=0.6]{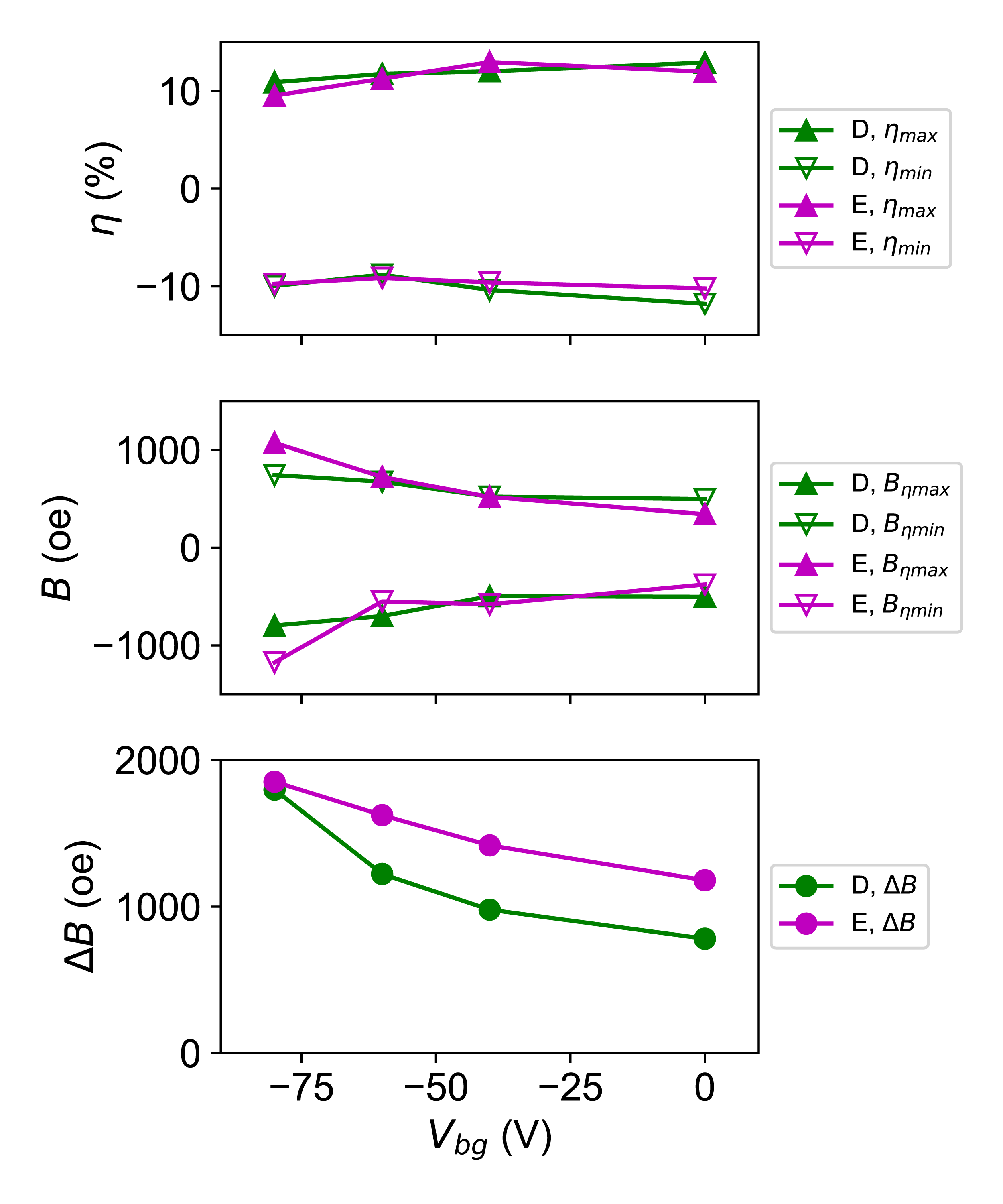}
\caption{Backgate dependence of diode efficiency of Devices D and E. Optimal diode efficiency $\eta_{\rm max(min)}$ (top) and corresponding optimal magnetic field $B_{\eta \rm min (\eta \rm max)}$ (middle) are plotted as a function of $V_{\rm bg}$ applied on the sample. The datapoints are extracted from main text Figure~3(f)(i) as well as Figure~\ref{D bg}(c)(f)(i) and Figure~\ref{E bg}(c)(f)(i). At $T=50$~mK, $I_c$ vs $B$ relation of Device~D or E resembles a M-shape, with $\Delta B$ splitting the two $I_c$ peaks. Here $\Delta B$ of Device~D and E are plotted as a function of $V_{\rm bg}$ in the bottom panel.}
\label{DE bg}
\end{figure}

\begin{figure}[p]
\centering
\includegraphics[scale=0.97]{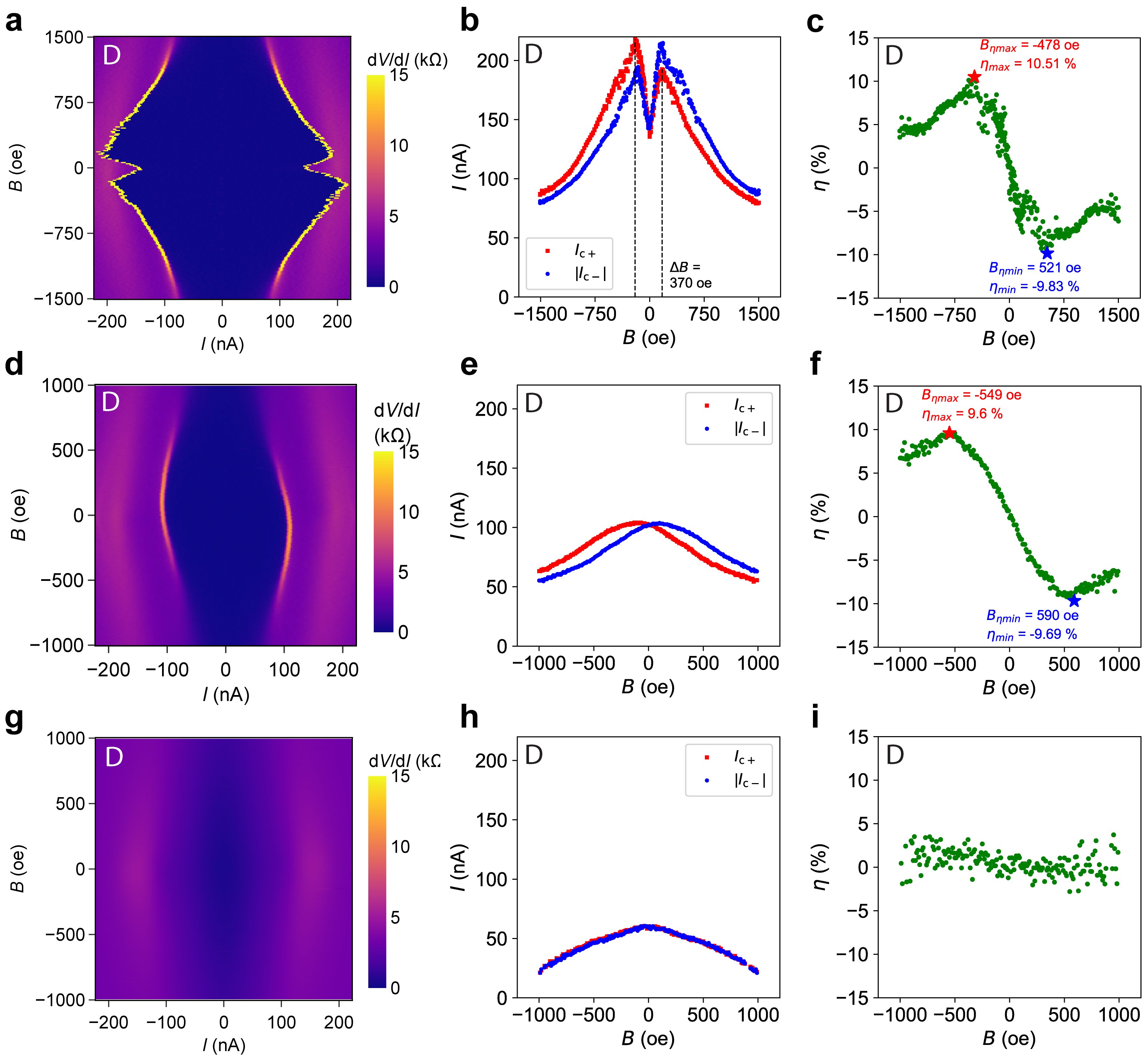}
\caption{Temperature dependence of Device~D. (a)(b)(c) $dV/dI$ vs $I$ vs $B$ intensity plot, $I_{c\pm}$ vs $B$ and $\eta$ vs $B$ relations of Device~D, measured at $T=500$~mK. The splitting $\Delta B$ between the two $I_c$ maxima is labeled, as well as the location of $\eta_{\rm max}$ and $\eta_{\rm min}$. (d)(e)(f) $dV/dI$ vs $I$ vs $B$ intensity plot, $I_{c\pm}$ vs $B$ and $\eta$ vs $B$ relations of Device~D, measured at $T=900$~mK. M-shaped $I_c$ vs $B$ feature is suppressed at this temperature. (g)(h)(i) $dV/dI$ vs $I$ vs $B$ intensity plot, $I_{c\pm}$ vs $B$ and $\eta$ vs $B$ relations of Device~D, measured at $T=1.2$~K. All plots in this figure were taken with backgate grounded ($V_{\rm bg}=0$~V).}
\label{D temp}
\end{figure}


\end{document}